\documentclass[10pt]{iopart}

\usepackage{graphicx}
\usepackage{dcolumn}
\usepackage{bm}
\usepackage{color}

\begin{document}

\title[Alfven Eigenmodes in DIII-D high poloidal $\beta$ discharges]{Analysis of Alfven Eigenmode destabilization in DIII-D high poloidal $\beta$ discharges using a Landau closure model}


\author{J. Varela}
\ead{rodriguezjv@ornl.gov}
\address{Oak Ridge National Laboratory, Oak Ridge, Tennessee 37831-8071, USA}
\author{D. A. Spong}
\address{Oak Ridge National Laboratory, Oak Ridge, Tennessee 37831-8071, USA}
\author{L. Garcia}
\address{Universidad Carlos III de Madrid, 28911 Leganes, Madrid, Spain}
\author{J. Huang}
\address{Institute of Plasma Physics, Chinese Academy of Science, Hefei, China}
\author{M. Murakami}
\address{Oak Ridge National Laboratory, Oak Ridge, Tennessee 37831-6069, USA}
\author{A. M. Garofalo}
\address{General Atomics, San Diego, CA, 92186, USA}
\author{J. P. Qian}
\address{Institute of Plasma Physics, Chinese Academy of sciences, Hefei 230031, China}
\author{C. T. Holcomb}
\address{Lawrence Livermore National Laboratory, Livermore, CA 94551}
\author{A. W. Hyatt}
\address{General Atomics, San Diego, CA, 92186, USA}
\author{J. R. Ferron}
\address{General Atomics, San Diego, CA, 92186, USA}
\author{C. S. Collins}
\address{General Atomics, San Diego, CA, 92186, USA}
\author{Q. L. Ren}
\address{Institute of Plasma Physics, Chinese Academy of sciences, Hefei 230031, China}
\author{J. McClenaghan}
\address{Oak Ridge Associated Universities}
\author{W. Guo}
\address{Institute of Plasma Physics, Chinese Academy of Science, Hefei, Anhui 230031, China}

\date{\today}

\begin{abstract}
Alfv\' en Eigenmodes are destabilized at the DIII-D pedestal during transient beta drops in high poloidal $\beta$ discharges with internal transport barriers (ITBs), driven by $n=1$ external kink modes, leading to energetic particle losses. There are two different scenarios in the thermal $\beta$ recovery phase: with bifurcation (two instability branches with different frequencies) or without bifurcation (single instability branch). We use the reduced MHD equations in a full 3D system, coupled with equations of density and parallel velocity moments for the energetic particles as well as the geodesic acoustic wave dynamics, to study the properties of the instabilities observed in the DIII-D high poloidal $\beta$ discharges and identify the conditions to trigger the bifurcation. The simulations suggest that instabilities with lower frequency in the bifurcation case are ballooning modes driven at the plasma pedestal, while the instability branch with higher frequencies are low n ($n<4$) Toroidal Alfv\' en Eigenmodes nearby the pedestal. The reverse shear region between the middle and plasma periphery in the non-bifurcated case avoids the excitation of ballooning modes at the pedestal, although Toroidal Alfv\' en Eigenmodes and Reverse Shear Alfv\' en Eigenmodes are unstable in the reverse shear region. The $n=1$ and $n=2$ Alfv\' en Eigenmode activity can be suppressed or minimized if the neutral beam injector (NBI) intensity is lower than the experimental value ($\beta_{f} < 0.03$). In addition, if the beam energy or neutral beam injector voltage is lower than in the experiment ($V_{th,f} / V_{A0} < 0.2$), the resonance between beam and thermal plasma is weaker. The $n=3,4,5$ and $6$ AE activity can't be fully suppressed, although the growth rate and frequency is smaller for an optimized neutral beam injector operation regime. In conclusion, AE activity in high poloidal $\beta$ discharges can be minimized for optimized NBI operation regimes.
\end{abstract}

%
%
%
%
%


\vspace{2pc}

This manuscript has been authored by UT-Battelle, LLC under Contract No. DE-AC05- 00OR22725 with the U.S. Department of Energy. The United States Government retains and the publisher, by accepting the article for publication, acknowledges that the United States Government retains a non-exclusive, paid-up, irrevocable, world-wide license to publish or reproduce the published form of this manuscript, or allow others to do so, for United States Government purposes. The Department of Energy will provide public access to these results of federally sponsored research in accordance with the DOE Public Access Plan (http://energy.gov/downloads/doe-public-access-plan).

\maketitle

\ioptwocol

\section{Introduction \label{sec:introduction}}

High poloidal $\beta$ discharges are a necessary component of tokamak steady state operation \cite{1,2,3,4,5,6}, based on bootstrap current and non inductive current drive \cite{7,11}. High poloidal $\beta$ discharges have smaller toroidal plasma currents leading to a reduced possibility of triggering plasma disruptions, improved MHD instability (second stability regime), favorable transport properties and higher confinement factor. In addition, the reactor extrapolation leads to a reasonable device size, fusion output power and a possible high $\beta_{p}$ ITER scenario. Fully non inductive high poloidal $\beta$ operations in DIII-D show large internal transport barriers (ITB) that improve the device performance ($H_{98} \ge 1.4$) as well as good confinement even for low rotation conditions, although there is a limit of $\beta_{p} \approx 1.9$ due to the Shafranov shift for turbulence suppression and and ITB formation.

High poloidal $\beta$ discharges in DIII-D with $\beta_{p} \approx 3$ and $q_{min}=3$ show a transient $\beta$, density, $q_{min}$ and rotation drop if the $n=1$ external kink mode is destabilized \cite{12}, leading to a disappearance of the ITB. Alfv\' en Eigenmode (AE) activity is enhanced after the onset of the external kink, inducing larger fast-ion transport losses and inhibiting or even preventing the $\beta$ recovery. Two scenarios were observed after the $\beta$ collapse: discharges with bifurcation (two instability branches with different frequencies driven in the middle plasma and at the pedestal) and without bifurcation (single instability branch). \textcolor{red}{The energetic particle losses driven by the destabilized AEs lead to a decrease of the expected neutron measurements up to a $50 \%$ in the bifurcation case and up to $60 \%$ in the non bifurcation case after the collapse of the thermal plasma $\beta$ \cite{12}. During the recovery phase of the thermal plasma $\beta$ the expected neutron rate remains between $20$ to $30 \%$ below the measurements before the collapse. For more information on these discharges please see reference \cite{12} where the numerical model results of the energetic particle losses are compared to the measurements, showing good agreement.}.

Energetic particle driven instabilities can enhance the transport of fusion produced alpha particles, energetic hydrogen neutral beams and particles heated using ion cyclotron resonance heating (ICRF) \cite{13,14,15}. The consequence is a decrease of the operation performance in devices  as TFTR, JET and DIII-D tokamaks or LHD and W7-AS stellarators \cite{16,17,18,19,20,21}. If the mode frequency resonates with the drift, bounce or transit frequencies of the energetic particles, the particle and diffusive losses increase. In addition, plasma instabilities as internal kinks \cite{22,23} or ballooning modes \cite{24} can be kinetically destabilized.

Alfv\' en Eigenmodes (AE) are driven in the spectral gaps of the shear Alfv\' en continua \cite{25,26}, destabilized by Super-Alfv\' enic alpha particles and energetic particles. Alfv\' en Eigenmode (AE) activity was observed before in several discharges and configurations \cite{27,28,29,30}. The different Alfv\' en eigenmode families ($n$ is the toroidal mode and $m$ the poloidal mode) are linked to frequency gaps produced by periodic variations of the Alfv\' en speed, for example: toroidicity induced Alfv\' en Eigenmodes (TAE) coupling $m$ with $m+1$ modes \cite{31,32,33}, beta induced Alfv\' en Eigenmodes driven by compressibility effects (BAE) \cite{34}, Reversed-shear Alfv\' en Eigenmodes (RSAE) due to local maxima/minima in the safety factor $q$ profile \cite{35}, Global Alfv\' en Eigenmodes (GAE) observed in the minimum of the Alfv\' en continua \cite{36,37}, ellipticity induced Alfv\' en Eigenmodes (EAE) coupling $m$ with $m+2$ modes \cite{38,39}, noncircularity induced Alfv\' en Eigenmodes (NAE) coupling $m$ with $m+3$ or higher \cite{40,41}.

DIII-D plasmas are heated by eight neutral beam injectors (NBI), six sources injected in the midplane (on axis) and 2 injected downwards at an angle (off axis). Six sources are injected in the direction of the plasma current (co-injected), including two tilted sources, and 2 source are injected opposite to the plasma current (counter-injected). The plasma is deuterium and the NBI also injects deuterium with a beam energy of $80$ keV ($2.25$ MW source). The destabilization of AE linked to strong NBI heating was measured before in DIII-D, triggering a large variety of AE instabilities as GAE \cite{42}, TAE \cite{43}, RSAE \cite{44}, BAE \cite{45}, EAE \cite{46} and NAE \cite{47}. The AE instabilities reduce the device performance, increasing the transport and enhancing energetic particle losses \cite{48,49,50}.

The aim of the present study is to analyze AE stability at the DIII-D pedestal during high poloidal $\beta$ discharges, comparing simulation results and experimental observations. We also study how the bifurcation scenario is originated, as well as the features of each instability frequency branch instabilities. In addition, we predict the optimal NBI operation regime to improve the plasma AE stability.

A set of simulations are performed using an updated version of the FAR3D code \cite{51,52,53}, adding the moment equations of the energetic ion density and parallel velocity \cite{54,55}. This numerical model, with the appropriate Landau closure relations, solves the reduced non-linear resistive MHD equations including the linear wave-particle resonance effects, required for Landau damping/growth, and the parallel momentum response of the thermal plasma, required for coupling to the geodesic acoustic waves \cite{35}. The code follows the evolution of six field variables, starting from equilibria calculated by the VMEC code \cite{56}. A methodology has been developed to calibrate Landau-closure models against more complete kinetic models and optimize the closure coefficients \cite{35}. The model includes Landau resonance couplings, but not fast ion FLR \cite{55} or Landau damping of the modes on the background ions/electrons \cite{54}. Methods for including these effects have been developed for the companion tokamak gyrofluid code TAEFL \cite{35}, and will be adapted to this 3D Landau fluid model as a topic for future research.​

This paper is organized as follows. The model equations, numerical scheme and equilibrium properties are described in section \ref{sec:model}. The results of the interpretation study for the case with bifurcation are in section \ref{sec:interpretation}. The results of the interpretation study for the case without bifurcation are in section \ref{sec:interpretation2}. The analysis of the optimized NBI operational regime is shown in section \ref{sec:optimization}. Finally, the conclusions of this paper are presented in section \ref{sec:conclusions}.

\section{Equations and numerical scheme \label{sec:model}}

For high-aspect ratio configurations with moderate $\beta$-values (of the order of the inverse aspect ratio), we can apply the method employed in Ref.\cite{57} for the derivation of the reduced set of equations, retaining the toroidal angle variation, to describe the evolution of the background plasma and fields. We obtain a reduced set of equations using the exact three-dimensional equilibrium. The effect of the energetic particle population is included in the formulation as moments of the kinetic equation truncated with a closure relation \cite{58}. These describe the evolution of the energetic particle density ($n_{f}$) and velocity moments parallel to the magnetic field lines ($v_{||f}$). The coefficients of the closure relation are selected to match a two-pole approximation of the plasma dispersion function.     

In the derivation of the reduced equations we assume high aspect ratio, medium $\beta$ (of the order of the inverse aspect ratio $\varepsilon=a/R_0$), small variation of the fields and small resistivity. The plasma velocity and perturbation of the magnetic field are defined as
\begin{equation}
 \mathbf{v} = \sqrt{g} R_0 \nabla \zeta \times \nabla \Phi, \quad\quad\quad  \mathbf{B} = R_0 \nabla \zeta \times \nabla \psi,
\end{equation}
where $\zeta$ is the toroidal angle, $\Phi$ is a stream function proportional to the electrostatic potential, and $\psi$ is the perturbation of the poloidal flux.

The equations, in dimensionless form, are
\begin{equation}
\frac{\partial \tilde \psi}{\partial t} =  \sqrt{g} B \nabla_\| \Phi  + \eta \varepsilon^2 J \tilde J^\zeta
\end{equation}
\begin{eqnarray} 
\frac{{\partial \tilde U}}{{\partial t}} =  -\epsilon v_{\zeta,eq} \frac{\partial U}{\partial \zeta} \nonumber\\
+ S^2 \left[{ \sqrt{g} B \nabla_\| J^\zeta - \frac{\beta_0}{2\varepsilon^2} \sqrt{g} \left( \nabla \sqrt{g} \times \nabla \tilde p \right)^\zeta }\right]   \nonumber\\
-  S^2 \left[{\frac{\beta_f}{2\varepsilon^2} \sqrt{g} \left( \nabla \sqrt{g} \times \nabla \tilde n_f \right)^\zeta }\right] 
\end{eqnarray} 
\begin{eqnarray}
\label{pressure}
\frac{\partial \tilde p}{\partial t} = -\epsilon v_{\zeta,eq} \frac{\partial p}{\partial \zeta} + \frac{dp_{eq}}{d\rho}\frac{1}{\rho}\frac{\partial \tilde \Phi}{\partial \theta} \nonumber\\
 +  \Gamma p_{eq}  \left[{ \sqrt{g} \left( \nabla \sqrt{g} \times \nabla \tilde \Phi \right)^\zeta - \nabla_\|  v_{\| th} }\right] 
\end{eqnarray} 
\begin{eqnarray}
\label{velthermal}
\frac{{\partial \tilde v_{\| th}}}{{\partial t}} = -\epsilon v_{\zeta,eq} \frac{\partial v_{||th}}{\partial \zeta} -  \frac{S^2 \beta_0}{n_{0,th}} \nabla_\| p 
\end{eqnarray}
\begin{eqnarray}
\label{nfast}
\frac{{\partial \tilde n_f}}{{\partial t}} = -\epsilon v_{\zeta,eq} \frac{\partial n_{f}}{\partial \zeta} - \frac{S  v_{th,f}^2}{\omega_{cy}}\ \Omega_d (\tilde n_f) - S  n_{f0} \nabla_\| v_{\| f}   \nonumber\\
- \varepsilon^2  n_{f0} \, \Omega_d (\tilde \Phi) + \varepsilon^2 n_{f0} \, \Omega_* (\tilde  \Phi) 
\end{eqnarray}
\begin{eqnarray}
\label{vfast}
\frac{{\partial \tilde v_{\| f}}}{{\partial t}} = -\epsilon v_{\zeta,eq} \frac{\partial v_{||f}}{\partial \zeta}  -  \frac{S  v_{th,f}^2}{\omega_{cy}} \, \Omega_d (\tilde v_{\| f}) \nonumber\\
- \left( \frac{\pi}{2} \right)^{1/2} S  v_{th,f} \left| \nabla_\|  v_{\| f}  \right| \nonumber\\
- \frac{S  v_{th,f}^2}{n_{f0}} \nabla_\| n_f + S \varepsilon^2  v_{th,f}^2 \, \Omega_* (\tilde \psi) 
\end{eqnarray}
Here, $U =  \sqrt g \left[{ \nabla  \times \left( {\rho _m \sqrt g {\bf{v}}} \right) }\right]^\zeta$ is the vorticity and $\rho_m$ the ion and electron mass density. The toroidal current density $J^{\zeta}$ is defined as:
\begin{eqnarray}
J^{\zeta} =  \frac{1}{\rho}\frac{\partial}{\partial \rho} \left(-\frac{g_{\rho\theta}}{\sqrt{g}}\frac{\partial \psi}{\partial \theta} + \rho \frac{g_{\theta\theta}}{\sqrt{g}}\frac{\partial \psi}{\partial \rho} \right) \nonumber\\
- \frac{1}{\rho} \frac{\partial}{\partial \theta} \left( \frac{g_{\rho\rho}}{\sqrt{g}}\frac{1}{\rho}\frac{\partial \psi}{\partial \theta} + \rho \frac{g_{\rho \theta}}{\sqrt{g}}\frac{\partial \psi}{\partial \rho} \right)
\end{eqnarray}
$v_{||th}$ is the parallel velocity of the thermal particles and $v_{\zeta,eq}$ is the equilibrium toroidal rotation. $n_{f}$ is normalized to the density at the magnetic axis $n_{f_{0}}$, $\Phi$ to $a^2B_{0}/\tau_{R}$ and $\Psi$ to $a^2B_{0}$. All lengths are normalized to a generalized minor radius $a$; the resistivity to $\eta_0$ (its value at the magnetic axis); the time to the resistive time $\tau_R = a^2 \mu_0 / \eta_0$; the magnetic field to $B_0$ (the averaged value at the magnetic axis); and the pressure to its equilibrium value at the magnetic axis. The Lundquist number $S$ is the ratio of the resistive time to the Alfv\' en time $\tau_{A0} = R_0 (\mu_0 \rho_m)^{1/2} / B_0$. $\rlap{-} \iota$ is the rotational transform, $v_{th,f} = \sqrt{T_{f}/m_{f}}$ the energetic particle thermal velocity normalized to the Alfv\' en velocity in the magnetic axis $v_{A0}$ and $\omega_{cy}$ the energetic particle cyclotron frequency times $\tau_{A0}$. $q_{f}$ is the charge, $T_{f}$ the temperature and $m_{f}$ the mass of the energetic particles. The $\Omega$ operators are defined as:
\begin{eqnarray}
\label{eq:omedrift}
\Omega_d = \frac{1}{2 B^4 \sqrt{g}}  \left[  \left( \frac{I}{\rho} \frac{\partial B^2}{\partial \zeta} - J \frac{1}{\rho} \frac{\partial B^2}{\partial \theta} \right) \frac{\partial}{\partial \rho}\right] \nonumber\\
-   \frac{1}{2 B^4 \sqrt{g}} \left[ \left( \rho \beta_* \frac{\partial B^2}{\partial \zeta} - J \frac{\partial B^2}{\partial \rho} \right) \frac{1}{\rho} \frac{\partial}{\partial \theta} \right] \nonumber\\ 
+ \frac{1}{2 B^4 \sqrt{g}} \left[ \left( \rho \beta_* \frac{1}{\rho} \frac{\partial B^2}{\partial \theta} -  \frac{I}{\rho} \frac{\partial B^2}{\partial \rho} \right) \frac{\partial}{\partial \zeta} \right]
\end{eqnarray}

\begin{eqnarray}
\label{eq:omestar}
\Omega_* = \frac{1}{B^2 \sqrt{g}} \frac{1}{n_{f0}} \frac{d n_{f0}}{d \rho} \left( \frac{I}{\rho} \frac{\partial}{\partial \zeta} - J \frac{1}{\rho} \frac{\partial}{\partial \theta} \right) 
\end{eqnarray}
Here the $\Omega_{d}$ operator is constructed to model the average drift velocity of a passing particle and $\Omega_{*}$ models its diamagnetic drift frequency. We also define the parallel gradient and curvature operators:
\begin{equation}
\label{eq:gradpar}
\nabla_\| f = \frac{1}{B \sqrt{g}} \left( \frac{\partial \tilde f}{\partial \zeta} +  \rlap{-} \iota \frac{\partial \tilde f}{\partial \theta} - \frac{\partial f_{eq}}{\partial \rho}  \frac{1}{\rho} \frac{\partial \tilde \psi}{\partial \theta} + \frac{1}{\rho} \frac{\partial f_{eq}}{\partial \theta} \frac{\partial \tilde \psi}{\partial \rho} \right)
\end{equation}
\begin{equation}
\label{eq:curv}
\sqrt{g} \left( \nabla \sqrt{g} \times \nabla \tilde f \right)^\zeta = \frac{\partial \sqrt{g} }{\partial \rho}  \frac{1}{\rho} \frac{\partial \tilde f}{\partial \theta} - \frac{1}{\rho} \frac{\partial \sqrt{g} }{\partial \theta} \frac{\partial \tilde f}{\partial \rho}
\end{equation}
with the Jacobian of the transformation:
\begin{equation}
\label{eq:Jac}
\frac{1}{\sqrt{g}} = \frac{B^2}{\varepsilon^2 (J+ \rlap{-} \iota I)}
\end{equation}

Equations~\ref{pressure} and~\ref{velthermal} introduce the parallel momentum response of the thermal plasma, required for coupling to the geodesic acoustic waves, accounting the geodesic compressibility in the frequency range of the geodesic acoustic mode (GAM) \cite{59,60}.

Equilibrium flux coordinates $(\rho, \theta, \zeta)$ are used. Here, $\rho$ is a generalized radial coordinate proportional to the square root of the toroidal flux function, and normalized to one at the edge. The flux coordinates used in the code are those described by Boozer \cite{61}, and $\sqrt g$ is the Jacobian of the coordinate transformation. All functions have equilibrium and perturbation components represented as: $ A = A_{eq} + \tilde{A} $. 

The FAR3D code uses finite differences in the radial direction and Fourier expansions in the two angular variables. The numerical scheme is semi-implicit in the linear terms. The nonlinear version uses a two semi-step method to ensure $(\Delta t)^2$ accuracy.

The present model was already used to study the AE activity in LHD \cite{62,63} and TJ-II \cite{64,65,66}, indicating reasonable agreement with the observations.

\subsection{Equilibrium properties}

We use fixed boundary results from the VMEC equilibrium code \cite{56} calculated using the DIII-D reconstruction of high poloidal $\beta$ discharges with bifurcation (shot 166495, case A) and without bifurcation (shot 166496, case B). The experimental constraints used in the equilibrium reconstruction are taken from magnetic data, MSE data, kinetic pressure and edge density profile from NEO model. Due to the fact that the FAR3D stability model is based on stellarator symmetry, we null out the up-down asymmetric terms in the VMEC shape and base the calculations of the current paper on up-down symmetric equilibria. Since the original experiments were run in single-null divertor mode, the equilibria we use here will be nearby, but slightly different from the experimental ones. We analyze three different phases during the 166495 shot: before the destabilization of the external kink ($t=3465$ ms, case A1), during the thermal $\beta$ collapse ($t=3585$ ms, case A2) and during the thermal $\beta$ recovery ($t=3650$ ms, case A3). We also analyze two phases during the 166496 shot:  before the destabilization of the external kink ($t=3345$ ms, case B1), and after the onset of the external kink ($t=3445$ ms, case B2). Table~\ref{Table:1} shows the main plasma properties.

\begin{table*}[t]
\centering
\begin{tabular}{c | c c c c c c c}
Case & $T_{i}(0)$ (keV) & $n_{i}(0)$ ($10^{20}$ m$^{-3}$) & $\beta_{th}(0)$ & $V_{A0}$ ($10^{6}$ m/s) & $n_{f}(0)$ ($10^{20}$ m$^{-3}$) & $\beta_{f}(0)$ & $\omega_{cy}\tau_{A0}$\\ \hline
A1 & 3.71 & 0.82 & 0.062 & 4.88 & 0.042 & 0.020 & 69.80 \\
A2 & 2.60 & 0.64 & 0.053 & 5.58 & 0.081 & 0.038 & 61.42 \\
A3 & 3.19 & 0.65 & 0.056 & 5.52 & 0.097 & 0.046 & 61.95 \\
B1 & 3.37 & 0.78 & 0.061 & 4.96 & 0.078 & 0.029 & 68.67 \\
B2 & 2.06 & 0.74 & 0.039 & 5.26 & 0.074 & 0.036 & 65.26 \\
\end{tabular}
\caption{Plasma properties (values at the magnetic axis). First column is the thermal ion temperature, second column is the thermal ion density, third column is the thermal $\beta$, forth column is the Alfv\' en velocity, fifth column is the energetic particle density, sixth column the energetic particle $\beta$ and seventh column the normalized cyclotron frequency.} \label{Table:1}
\end{table*}

\begin{figure*}[h!]
\centering
\includegraphics[width=0.8\textwidth]{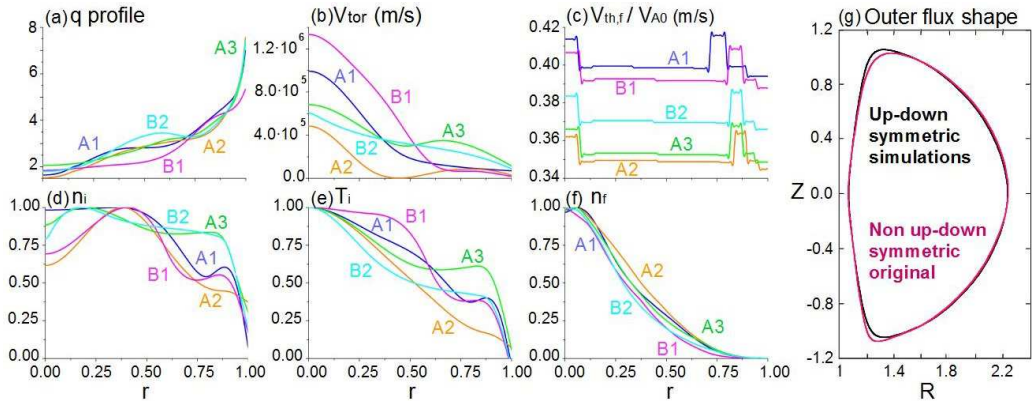}
\caption{(a) q profile, (b) toroidal rotation, (c) $V_{th,f}/V_{A0}$ ratio, (d) normalized thermal ion density, (e) normalized thermal ion temperature, (f) normalized energetic particle density and (g) outer flux shape. }\label{FIG:1}
\end{figure*}

The magnetic field at the magnetic axis is $2$ T and the averaged inverse aspect ratio is $\varepsilon=0.47$. The energy of the injected particles by the NBI is $T_{i}(0) = 49.32$ keV ($v_{th,f} = 2.173 \cdot 10^{6}$ m/s). For each configuration, Figure~\ref{FIG:1} panel (a) shows the q profile, panel (b) the toroidal rotation, panel (c) the $V_{th,f}/V_{A0}$ ratio, panel (d) the normalized thermal ion density, panel (e) the normalized thermal ion temperature, panel (f) the normalized energetic particle density and panel (g) the outer flux shape. We use in the simulations an up-down symmetric equilibria (black line) similar to the original case (purple line).

\begin{figure}[h!]
\centering
\includegraphics[width=0.5\textwidth]{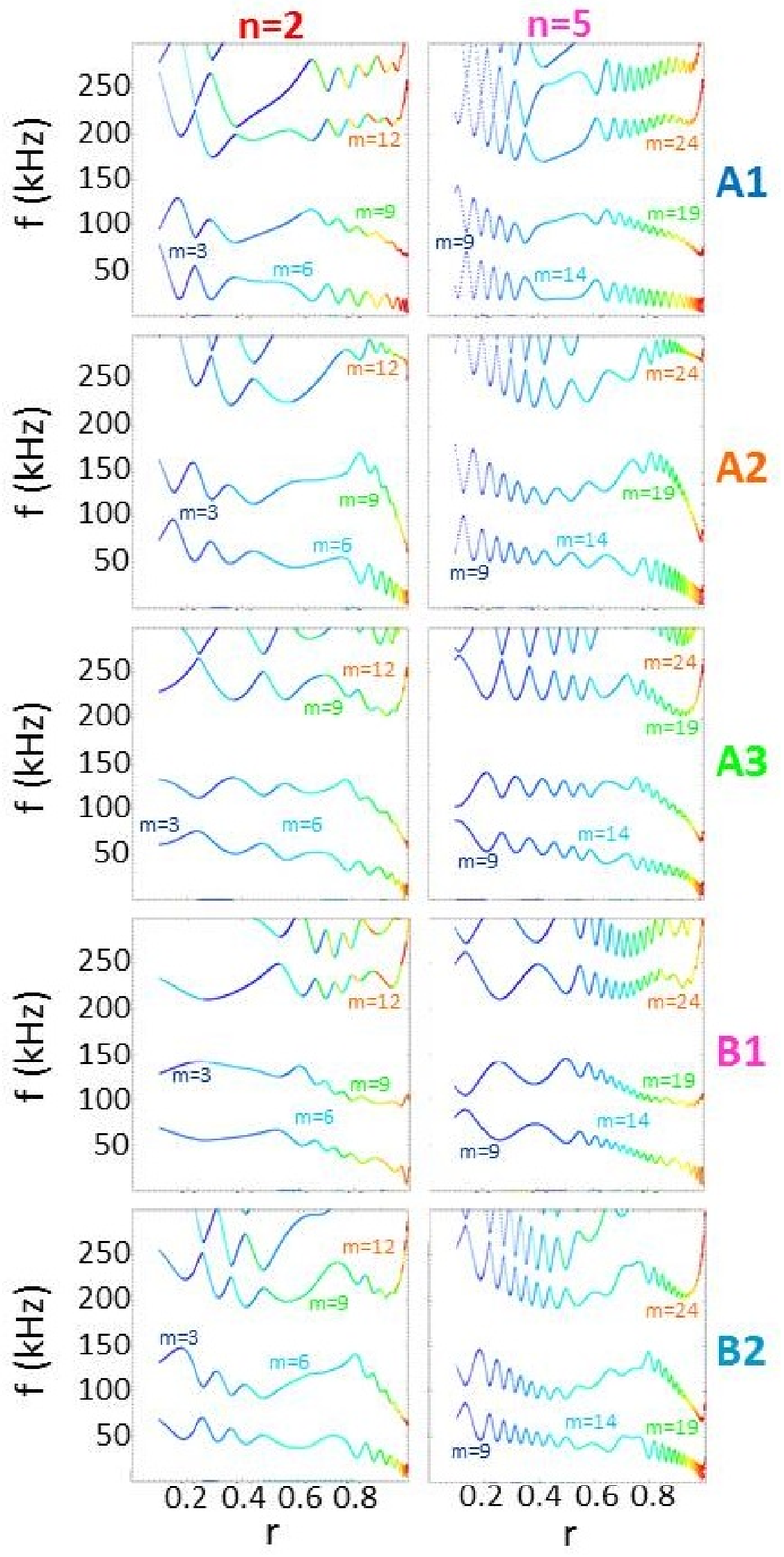}
\caption{Alfv\' en gaps of the $n=2$ and $5$ toroidal modes for the cases A1, A2, A3, B1 and B2.}\label{FIG:2}
\end{figure}

Figure~\ref{FIG:2} shows the Alfv\' en gaps of $n=2$ and $5$ toroidal modes for the cases A1, A2, A3, B1 and B2. There are four main Alfv\' en gaps: below $30$ kHz, around $75$, $180$ and $250$ kHz. TAEs frequencies are destabilized in between the $[75,180]$ kHz gaps, EAEs between the $[180,250]$ kHz gaps, NAE if $f>250$ Khz. BAE, BAAE and GAE are destabilized below $f= 75$ kHz gap.

\subsection{Simulations parameters}

The simulations are performed with a uniform radial grid of 1000 points. The dynamic and equilibrium toroidal (n) and poloidal (m) modes included in the study are summarized in table~\ref{Table:2} for each case. There are two different dynamic mode selections for cases A3 and B2, namely A3b and B2b, required to analyze the AE stability at the plasma pedestal of case A3 ($\rho > 0.9$) and in the reverse shear region of case B2 ($\rho = [0.6 , 0.85]$). The plasma core is not included in the analysis ($\rho < 0.5$) if the dominant mode of the simulations is a MHD instability nearby the magnetic axis (interchange mode), because the aim of the study is to analyze the AE activity at the pedestal. The equilibrium mode selection is the same for all cases. In the following, the mode number notation is $m/n$, consistent with the $q=m/n$ definition for the associated resonance.

\begin{table*}[h]
\centering
\begin{tabular}{c | c c c c c c c}
\hline
(n) & A1 (m) & A2 (m) & A3 (m) & A3b (m) & B1 (m) & B2 (m) & B2b (m) \\ \hline
$1$ & $[1,3]$ & $[3,7]$ & $[3,6]$ & -- & $[1,4]$ & $[4,6]$ & -- \\
$2$ & $[5,12]$ & $[6,12]$ & $[6,9]$ & -- & $[3,12]$ & $[4,13]$ & $[6,9]$ \\
$3$ & $[8,15]$ & $[9,18]$ & $[8,15]$ & -- & $[5,17]$ & $[6,14]$ & $[9,12]$ \\
$4$ & $[10,18]$ & $[12,20]$ & $[13,18]$ & $[14,19]$ & $[7,20]$ & $[8,18]$ & $[13,16]$ \\
$5$ & $[13,21]$ & $[15,24]$ & $[16,21]$ & $[18,24]$ & $[9,22]$ & $[9,19]$ & $[16,21]$ \\
$6$ & $[14,24]$ & $[18,27]$ & $[18,22]$ & $[23,29]$ & $[11,24]$ & $[10,20]$ & $[19,25]$ \\ \hline
(n) & & & & All cases (m) & & & \\ \hline
$0$ & & & & $[0,9]$ & & & \\ \hline
\end{tabular}
\caption{Dynamic and equilibrium toroidal (n) and poloidal (m) modes.} \label{Table:2}
\end{table*}

The kinetic closure moment equations (6) and (7) break the usual MHD parities. This is taken into account by including both parities $sin(m\theta + n\zeta)$ and $cos(m\theta + n\zeta)$ for all dynamic variables, and allowing for both a growth rate and real frequency in the eigenmode time series analysis. The convention of the code is, in case of the pressure eigenfunction, that $n > 0$ corresponds to the Fourier component $\cos(m\theta + n\zeta)$ and $n < 0$ to $\sin(-m\theta - n\zeta)$. For example, the Fourier component for mode $-7/2$ is $\cos(-7\theta + 2\zeta)$ and for the mode $7/-2$ is $\sin(-7\theta + 2\zeta)$. The magnetic Lundquist number is $S=5\cdot 10^6$ similar to the experimental value in the middle of the plasma.

The density ratio between the energetic particles and bulk plasma ($n_{f}(0)/n_{e}(0)$) at the magnetic axis is controlled through the $\beta_{f}=$ value, linked to the NBI injection intensity, calculated by the code TRANSP without the effect of the anomalous beam ion transport. \textcolor{red}{Including anomalous transport effects should lead to an enhancement of the energetic particle losses, modifying the density and temperature profiles of the energetic particles used in the model. Such effects could steepen the fast ion gradients near the edge and would mostly influence the growth rates for the modes and not the real frequencies. Models for such corrections are currently in the developed and testing phase \cite{67}; for this reason, we neglect this correction for now, considering the simulation results as a first order approximation.}

The ratio between the energetic particle thermal velocity and Alfv\' en velocity at the magnetic axis ($v_{th,f}/v_{A0}$) controls the resonance efficiency between AE and energetic particles, associated with the NBI voltage or beam energy. \textcolor{red}{The Landau closure model used here is based on two moment equations for the fast ions which is equivalent to a two-pole approximation to the plasma dispersion relation. This translates to a Lorentzian energy distribution function; to lowest order the Lorentzian can be matched either to a Maxwellian or a slowing-down distribution by choosing an equivalent average energy. For the results given in this paper, we have matched to the mean energy of a slowing-down distribution function. A more precise matching to the resonance function for a slowing-down distribution can be obtained by including higher moment equations for the fast ions. FAR3D has recently been extended to three and four moment versions, for improved matching to a variety of non-Maxellian distributions. These require further testing and calibration, using the methods presented in \cite{35} and will be the topic of future research.}

\section{Interpretation study of the bifurcation case \label{sec:interpretation}}

\begin{figure*}[h!]
\centering
\includegraphics[width=0.8\textwidth]{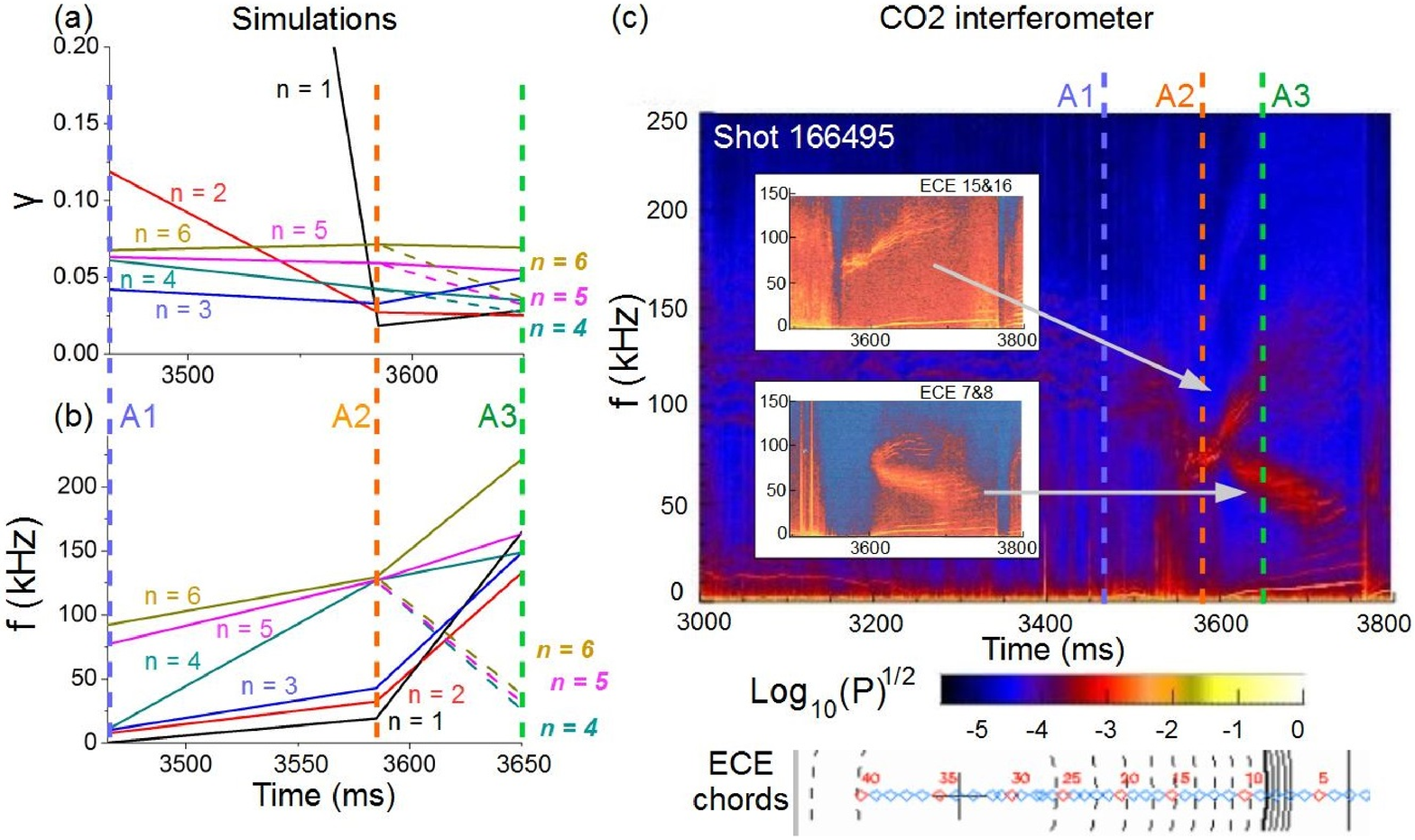}
\caption{Instabilities growth rate (a) and frequency (b) in the bifurcation case at different discharge phases: before the collapse (A1), during the collapse (A2) and after the collapse (A3). The solid lines indicate the simulations including all the modes and the dashed lines the simulations limited to the pedestal. The solid italic symbols indicate the mode number of the simulations limited to the pedestal. Panel (c) shows the CO2 interferometer data (sub-panels indicate the ECE data at different chords and the gray arrows the instability analyzed).}\label{FIG:3}
\end{figure*}

This section shows the interpretation study performed to analyze high poloidal $\beta$ configurations with bifurcation at different discharge phases: before, during and after the collapse. Figure~\ref{FIG:3} shows the growth rate ($\gamma$), panel a, and frequency ($f$), panel b, of the instabilities driven in each discharge phase (panel C, CO2 interferometer and ECE data). Before the collapse the plasma is AE unstable for all modes except $n=1$. The frequency of $n=2,3,4$ instabilities is $[15 - 20]$ kHz, $75$ kHz for $n=5$ and $90$ kHz for $n=6$, compatible with the observations. The $n=1$ is an MHD instability (interchange mode), showing the largest growth rate followed by $n=2$. During the collapse phase, $n=1$ is also AE unstable, with a frequency of $20$ kHz. The other mode frequencies increase too, particularly $n=4$, reaching the same frequency of $n=5$ and $6$ instabilities, around $125$ kHz. High $n$ instabilities have the largest growth rates. After the collapse, the instability frequencies further increase up to $[125 - 225]$ kHz, although if the simulations are limited to the plasma pedestal, the instability frequency drops for high $n$ modes to $[25-40]$ kHz, consistent with the two instability branches observed in the bifurcation case. Consequently, the low frequency instability branch is linked to the destabilization of the pedestal region (large $n$), while the large frequency instability branch is AE activity driven between the middle plasma and the pedestal. A detailed analysis of the instabilities driven in each discharge phase is done in the Appendix.

\begin{figure}[h!]
\centering
\includegraphics[width=0.5\textwidth]{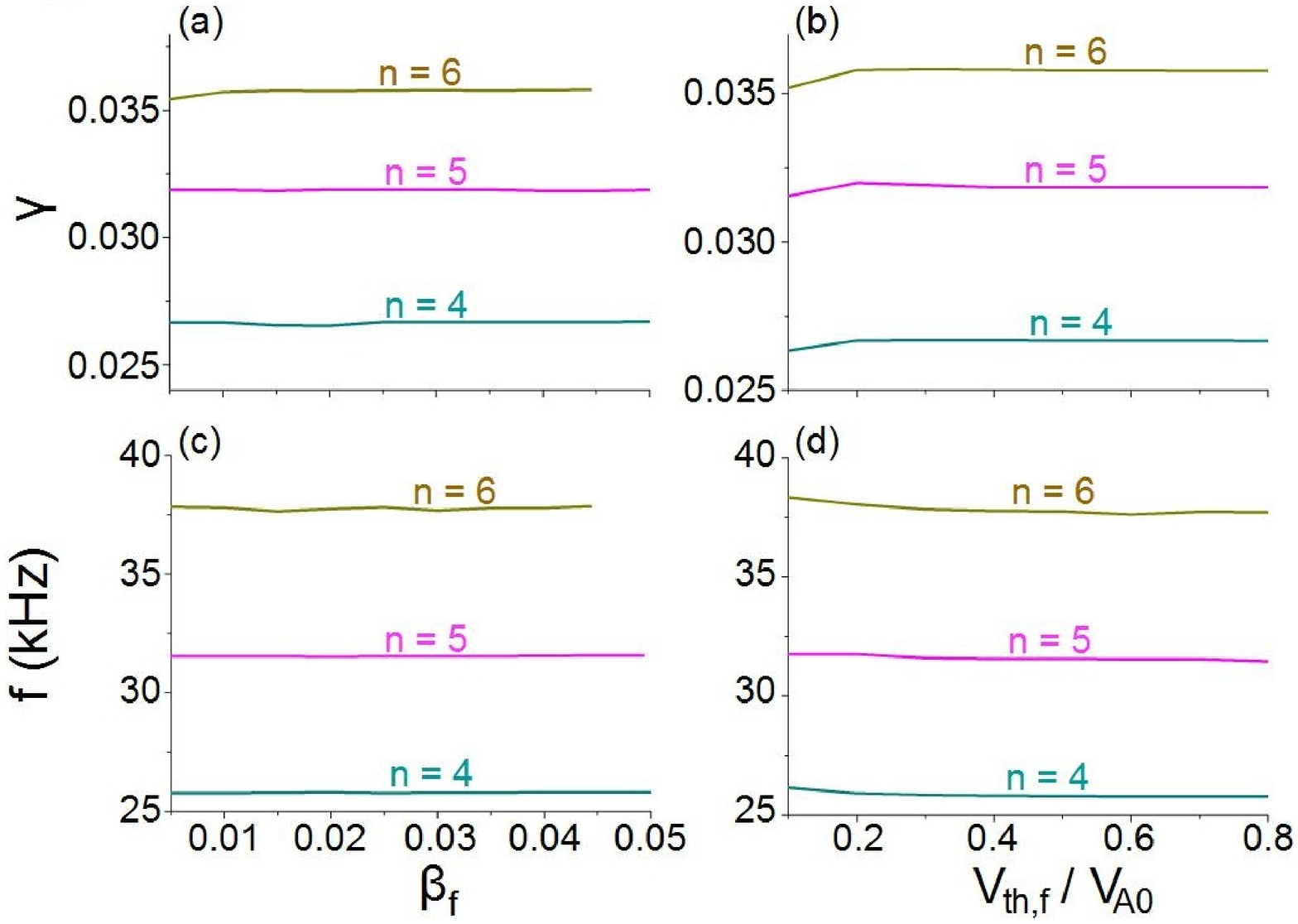}
\caption{Instability growth rate (a) and frequency (c) dependency with $\beta_{f}$. Instability growth rate (b) and frequency (d) dependency with $V_{th,f}/V_{A0}$ ratio.}\label{FIG:4}
\end{figure}

Figure~\ref{FIG:4} shows the effect of $\beta_{f}$ (NBI injection intensity) and $V_{th,f}/V_{A0}$ ratios (NBI voltage or energetic particle energy) on the instability growth rates and frequencies at the plasma pedestal. No dependency is observed with $\beta_{f}$, panel (a) and (c), or $V_{th,f}/V_{A0}$, panel (b) and (d), pointing out that it is an MHD instability (ballooning mode).

\begin{figure}[h!]
\centering
\includegraphics[width=0.5\textwidth]{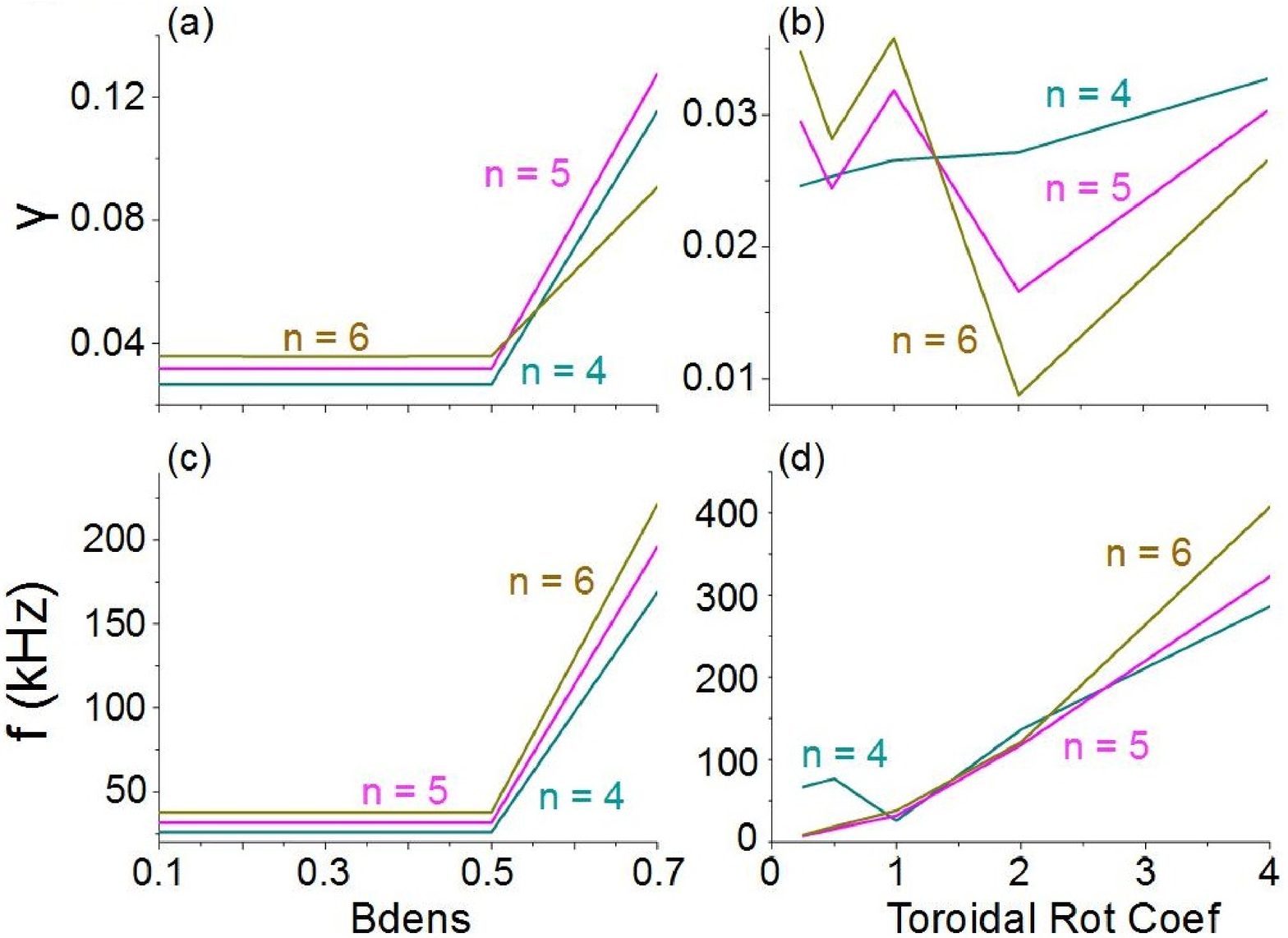}
\caption{Instability growth rate (a) and frequency (c) dependency with the location of the energetic particle density gradient. Instability growth rate (b) and frequency (d) dependency with the toroidal rotation.}\label{FIG:5}
\end{figure}

Figure~\ref{FIG:5} shows the effect of the energetic particle density gradient location, panels (a) and (c), and the toroidal rotation, panels (b) and (d), on the instability growth rates and frequencies at the plasma pedestal. The AE are only destabilized if the energetic particle density gradient is located in the plasma periphery ($\rho = 0.7$), leading to a large increase of the instability growth rate and frequency, and the modes frequency increase with the toroidal rotation. The effect of the energetic particle density gradient location and toroidal rotation on the instabilities is shown in the Appendix.

In summary, the bifurcation observed in shot 166495 is caused by the destabilization of low $n$ AE activity between the middle plasma and the periphery (high frequency branch) and ballooning modes at the plasma pedestal (low frequency branch). The equilibria after the collapse shows the larger thermal ion density and temperature gradient, see Figure~\ref{FIG:1}. In addition, there is a local maximum of the plasma toroidal rotation nearby the pedestal. Such equilibrium features lead to the decoupling of the plasma pedestal region ($\rho>0.9$) from the middle-periphery plasma region, together with the weak effect of localized ballooning modes on the rest of the plasma. Consequently, the coupling of poloidal modes of the same toroidal family between the pedestal and the rest of the plasma is weak, so the ballooning modes triggered at the pedestal can evolve independently, creating the instability frequency bifurcation.

\section{Interpretation study: non bifurcation case \label{sec:interpretation2}}

\begin{figure*}[h!]
\centering
\includegraphics[width=0.8\textwidth]{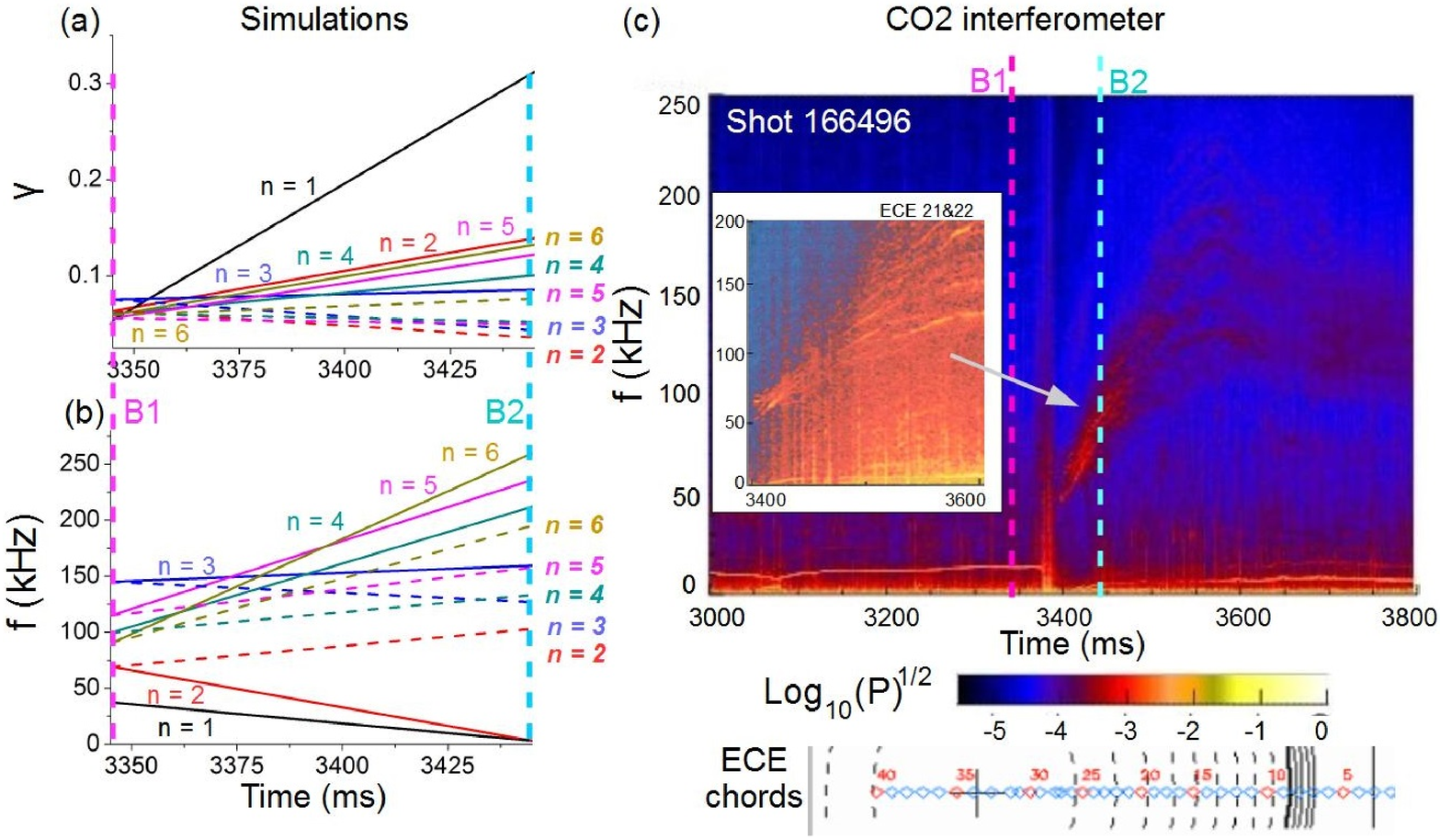}
\caption{Instabilities growth rate (a) and frequency (b) in the non bifurcation case at different discharge phases: before the collapse (B1) and after the collapse (B2). The solid lines indicate the simulations including all the modes and the dashed lines the simulations limited to the pedestal. The solid italic symbols indicate the mode number of the simulations limited to the reverse shear region. Panel (c) shows the CO2 interferometer data (sub-panels indicate the ECE data at different chords and the gray arrow the instability analyzed).}\label{FIG:6}
\end{figure*}

In this section the interpretation study of high poloidal $\beta$ configurations without bifurcation at different discharge phases is performed: before and after the collapse. Figure~\ref{FIG:6} indicates the growth rate (panel A) and frequency (panel B) of the instabilities driven in each discharge phase (panel C, CO2 interferometer and ECE data). In the phase before the collapse (B1) the plasma is AE unstable for all modes with frequencies in the range of $[40 - 145]$ kHz. The $n=3$ instability shows the largest growth rate. After the collapse (B2), the modes $n=1$ and $2$ are AE stable (interchange modes), $n=3$ frequency is almost the same and the frequency of $n=4,5,6$ instabilities increases. The range of frequencies for $n=4,5,6$ instabilities is $[200 - 250]$ kHz, larger than the observations (around $[90 - 150]$ kHz). Such discrepancy disappears if we perform simulations limited to the reverse shear region, pointing out that the measured instability should be AE activity driven at the reverse shear region. 

\begin{figure}[h!]
\centering
\includegraphics[width=0.5\textwidth]{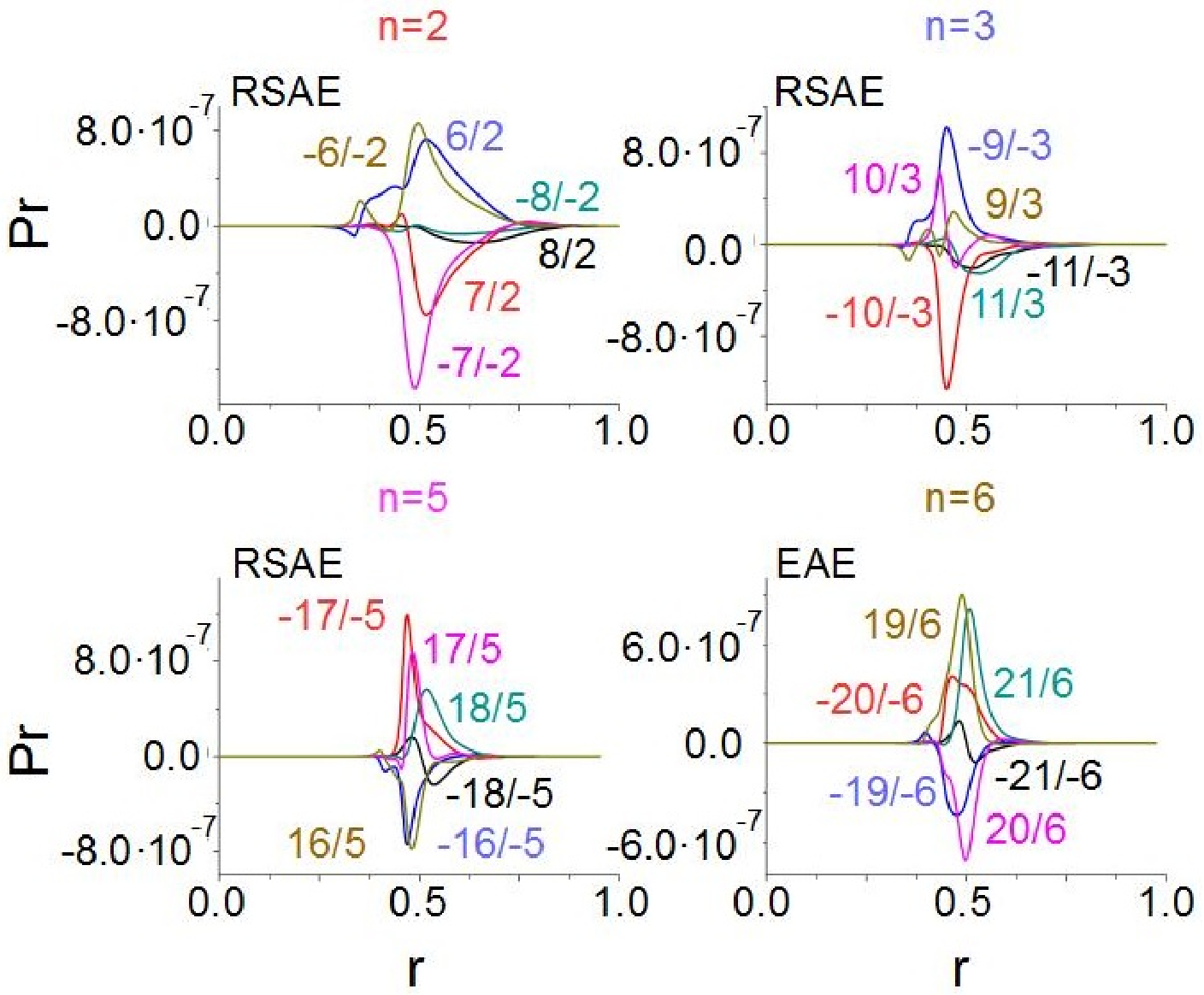}
\caption{Pressure eigenfunctions of $n=1,2,4,5$ instabilities in the case without bifurcation for the discharge phase after the collapse (simulations limited to the reverse shear region). Each panel includes the instability type.}\label{FIG:7}
\end{figure}

\begin{figure}[h!]
\centering
\includegraphics[width=0.3\textwidth]{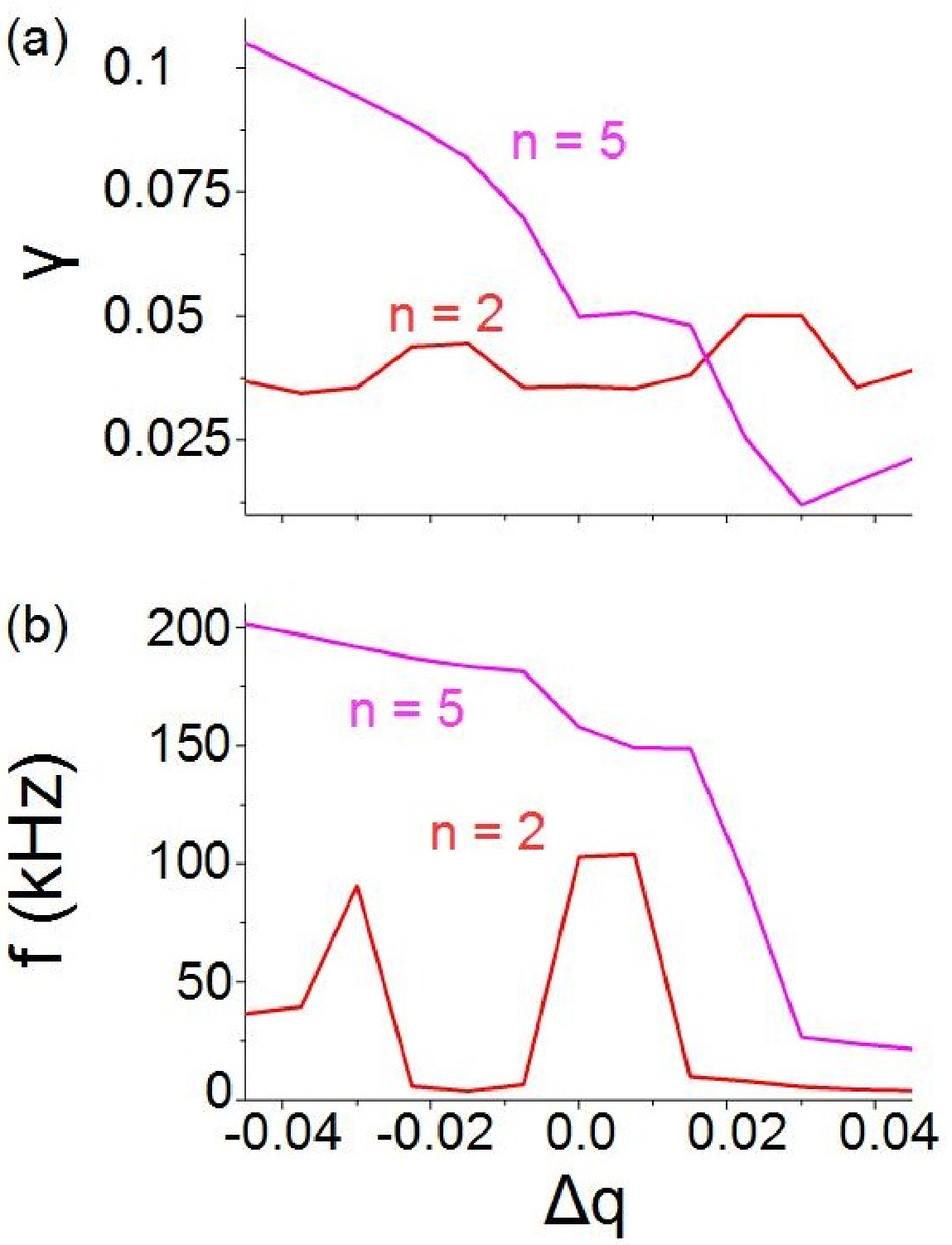}
\caption{Instability growth rate (a) and frequency (b) dependency with the iota profile displaced by $q=q_{0} \pm \Delta q$ with $\Delta q = [-0.045,0.045]$ each $0.0075$. Case B2b.}\label{FIG:8}
\end{figure}

Figure~\ref{FIG:7} shows the pressure eigenfunction of $n=1,3,5,6$ instabilities in the phase after the collapse for simulations limited to the reverse shear region (B2b). The $n=2,3,4,5$ instabilities are triggered at the beginning of the reverse shear region. The growth rate and frequency of $n=2,3,4,5$ instabilities sweep if the iota profile is displaced ($q=q_{0} \pm \Delta q$ with $\Delta q = [-0.045,0.045]$ each $0.0075$), Figure~\ref{FIG:8}. This is a feature of the RSAE consistent with the instability frequency sweeping measured after the collapse. The $n=6$ instability shows a larger toroidal mode couplings so it is a EAE.

In summary, the shot 166496 doesn't show a bifurcation because the toroidal mode coupling is stronger compared to the bifurcation case, and the evolution of the modes in the pedestal is linked to the modes in the reverse shear region. In addition, no local maxima of the toroidal rotation in the plasma periphery are observed. Consequently, localized modes as ballooning modes are stable although global modes as RSAE and TAE/EAE are triggered in the reverse shear region. RSAE and TAE/EAE frequencies are in the range of $[90 - 150]$ kHz, consistent with the observations that show an extended phase of frequency sweeping after the collapse, linked to the destabilization of the RSAE, not observed in the bifurcation case.

The next stage of the analysis is dedicated to the identification of the optimal NBI operational regime to suppress or minimize the negative effects of the AE activity on the plasma stability and transport. 

\section{Optimization of the NBI operation regime in high poloidal $\beta$ discharges \label{sec:optimization}}

This section shows the results of the parametric study performed to optimize the NBI operation in the discharge with bifurcation (during the collapse and after the collapse) and without bifurcation (after the collapse). In the analysis we identify the optimal NBI injection intensity ($\beta_{f}$) to avoid the destabilization of AEs, as well as the NBI voltage ($V_{th,f}/V_{A0}$) to operate in the weak resonant regime.

\subsection{Bifurcation case}

\begin{figure}[h!]
\centering
\includegraphics[width=0.5\textwidth]{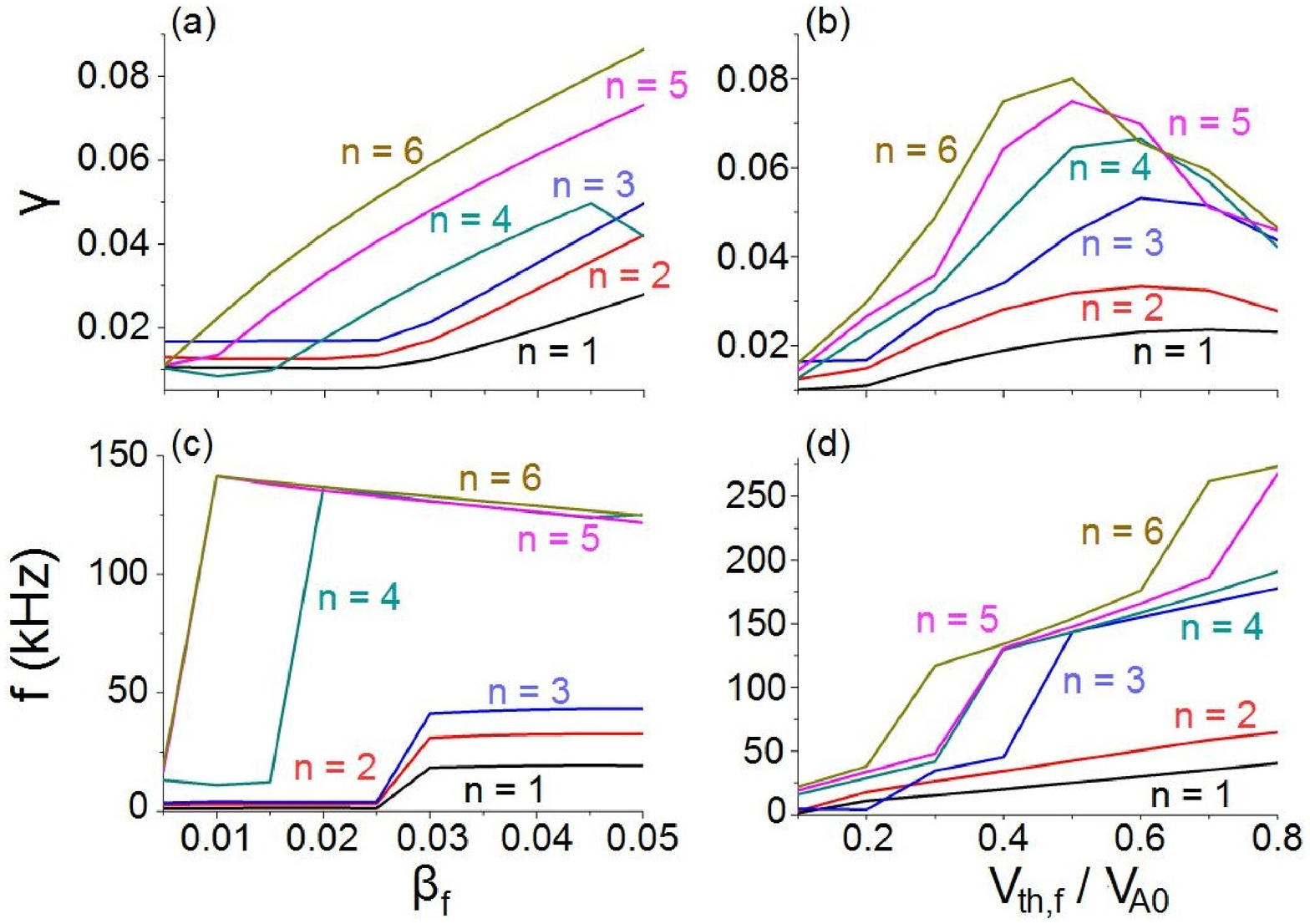}
\caption{Instability growth rate (a) and frequency (c) dependency with $\beta_{f}$. Instability growth rate (b) and frequency (d) dependency with  $V_{th,f}/V_{A0}$ ratio. Case A2.}\label{FIG:9}
\end{figure}

Figure~\ref{FIG:9} shows the effect of $\beta_{f}$ and the $V_{th,f}/V_{A0}$ ratio on the instability growth rate and frequency during the collapse. In the A2 phase, the NBI experimental operation regime is $\beta_{f} \approx 0.038$ and $V_{th,f}/V_{A0} \approx 0.35$, so all modes are AE unstable. The $n=5,6$ AE are destabilized if $\beta_{f} > 0.005$, $n=4$ AE if $\beta_{f} > 0.02$ and $n=1,2,3$ AE if $\beta_{f} > 0.03$, panels (a) and (c). The modes with larger growth rate are $n=5,6$. The NBI operates near the strong resonance regime, between $[0.4 - 0.6]$, panels (b) and (d). Increasing the beam energy leads to instabilities with higher frequencies. Optimization of the NBI operation requires lower beam energies, $V_{th,f}/V_{A0} < 0.2$, to operate in the weak resonance regime, leading to the stabilization of the $n=1$ AE. In addition, if the NBI injection intensity is weaker, $\beta_{f} < 0.03$, $n=2,3$ instabilities are AE stable.

\begin{figure}[h!]
\centering
\includegraphics[width=0.5\textwidth]{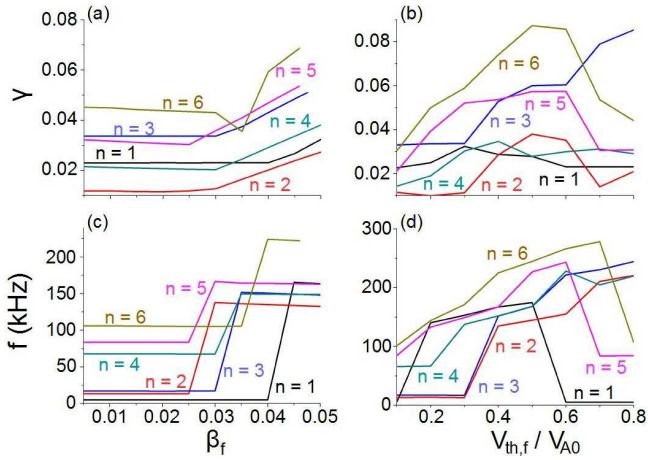}
\caption{Instability growth rate (a) and frequency (c) dependency with $\beta_{f}$. Instability growth rate (b) and frequency (d) dependency with $V_{th,f}/V_{A0}$ ratio. Case A3.}\label{FIG:10}
\end{figure}

Figure~\ref{FIG:10} analyzes the effect of $\beta_{f}$ and the $V_{th,f}/V_{A0}$ ratio on the instability growth rate and frequency after the collapse. The critical $\beta_{f}$ to destabilize the $n=1$ AE is $0.045$, $0.03$ for $n=2$ and $0.035$ for $n=3$, panels (a) and (c). The $n=4,5,6$ are AE unstable for all $\beta_{f}$ values. There are also critical $\beta_{f}$ values to destabilize $n=4,5,6$ AE with higher frequency and growth rate. Such critical $\beta_{f}$ is $0.035$ for $n=4$, $0.03$ for $n=5$ and $0.04$ for $n=6$. The NBI injection intensity in the A3 case is $\beta_{f} \approx 0.046$, so all modes are AE unstable. If the NBI injection intensity is weaker, $\beta_{f} < 0.03$, only $n=4,5,6$ modes are AE unstable, showing lower frequencies and growth rates compared to the experiment. The $n=1$ AE is destabilized if $V_{th,f}/V_{A0} = [0.2 - 0.5] $. The $n=2$ and $3$ are AE stable if $V_{th,f}/V_{A0} < 0.3$. The $n=4$ AE is unstable for all $V_{th,f}/V_{A0}$ ratios, although the instability frequency and growth rate decrease for ratios smaller than 0.2. The modes $n=5$ and $6$ are also AE unstable but the frequency and growth rate drop as the beam energy decreases. The NBI operational regime is $V_{th,f}/V_{A0} \approx 0.37$, close to the strong resonance regime, therefore the NBI operation optimization requires a $V_{th,f}/V_{A0} < 0.3$, to keep $n=2$ and $3$ AE stable and the growth rate and frequency of the $n=4,5,6$ AE small.

\begin{figure}[h!]
\centering
\includegraphics[width=0.5\textwidth]{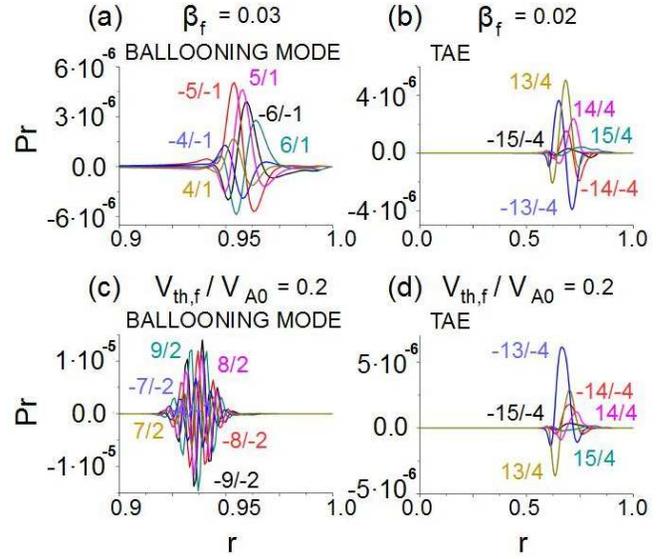}
\caption{Pressure eigenfunction of (a) $n=1$ instability for $\beta_{f}=0.03$, (b) $n=4$ instability for $\beta_{f}=0.02$, (c) $n=2$ and (d) $n=4$ instability for $V_{th,f}/V_{A0}=0.2$. Case A3. Each panel includes the instability type.}\label{FIG:11}
\end{figure}

Figure~\ref{FIG:11} shows the pressure eigenfunction of the $n=1,2,4$ instabilities for different $\beta_{f}$ and $V_{th,f}/V_{A0}$ ratios after the collapse. Compared to the eigenfunctions analyzed in Figure~\ref{FIG:3}, the $n=1$ instability for $\beta_{f}=0.03$ and $n=2$ for $V_{th,f}/V_{A0}=0.2$ are ballooning modes, not TAE, panels (a) and (c). If $\beta_{f}=0.02$ or $V_{th,f}/V_{A0}=0.2$, the $n=4$ instability is still a TAE although the toroidal coupling is enhanced. Consequently, the $n=1,2$ AE are stable and the frequency and growth rate of the $n=4$ AE is minimized if the NBI injection intensity and voltage are smaller than in the experiment.

\subsection{Non bifurcation case}

\begin{figure}[h!]
\centering
\includegraphics[width=0.5\textwidth]{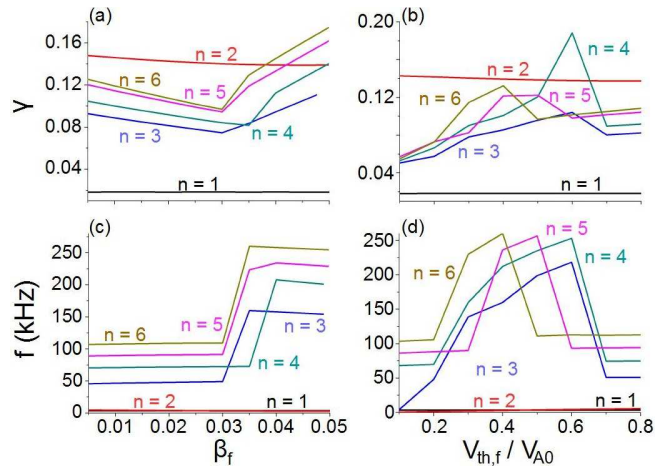}
\caption{Instability growth rate (a) and frequency (c) dependency with $\beta_{f}$. Instability growth rate (b) and frequency (d) dependency with $V_{th,f}/V_{A0}$ ratio. Case B2.}\label{FIG:12}
\end{figure}

Figure~\ref{FIG:12} analyzes the effect of the $\beta_{f}$ and $V_{th,f}/V_{A0}$ ratio on instability growth rates and frequencies after the collapse. Modes $n=1$ and $2$ are AE stable. Modes $ n=3,4,5,6$ are AE unstable for all $\beta_{f}$ values, as shown in panels (a) and (c). There is a critical $\beta_{f}$ value to destabilize  $ n=3,4,5,6$ AE instabilities with larger frequency and growth rates $\beta_{f} = 0.035$ for $n=3,5,6$ and $0.04$ for $n=4$. The experimental NBI injection intensity is $\beta_{f} \approx 0.036$ so all modes are AE unstable. In addition, the $n=3,5,6$ AE triggered have large frequency and growth rate. NBI optimal operation requires lower injection intensity, $\beta_{f} < 0.035$. Modes $n=1,2$ are MHD instabilities for all $V_{th,f}/V_{A0}$ ratios, panels (b) and (d). Between $V_{th,f}/V_{A0} = [0.3-0.6]$, $n= 3,4,5,6$ modes are in the strong resonance regime. In the experiment $V_{th,f}/V_{A0} \approx 0.37$, inside the strong resonance regime. Optimal NBI operation requires a lower NBI voltage, $V_{th,f}/V_{A0} < 0.3$, minimizing the $n=3,4,5,6$ AE frequency and growth rate.

\begin{figure}[h!]
\centering
\includegraphics[width=0.5\textwidth]{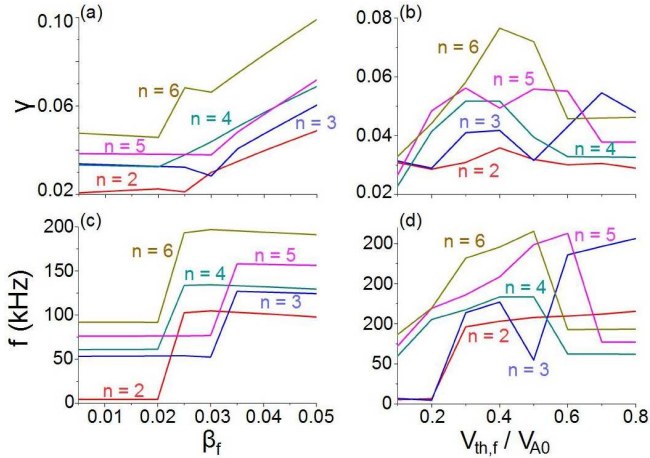}
\caption{Instability growth rate (a) and frequency (c) dependency with $\beta_{f}$. Instability growth rate (b) and frequency (d) dependency with $V_{th,f}/V_{A0}$ ratio. Case B2b.}\label{FIG:13}
\end{figure}

Figure~\ref{FIG:13} analyzes the effect of $\beta_{f}$ and $V_{th,f}/V_{A0}$ ratios on the instability growth rate and frequency after the collapse for simulations limited to the reverse shear region. The critical $\beta_{f}$ to destabilize the $n=2$ AE is $0.025$, panel (a) and (c). The $n=3,4,5,6$ are AE unstable for all $\beta_{f}$ values, although there is a critical $\beta_{f}$ to trigger AE instabilities with larger frequency and growth rate: $0.025$ for $n=4,6$ and $0.035$ for $n=3,5$. All modes are AE unstable in the experiment and $n=3,4,5,6$ AE show large frequencies and growth rates. Optimal NBI operation requires lower injection intensity, $\beta_{f} < 0.02$, and NBI voltage, $V_{th,f}/V_{A0} < 0.3$.

\begin{figure}[h!]
\centering
\includegraphics[width=0.5\textwidth]{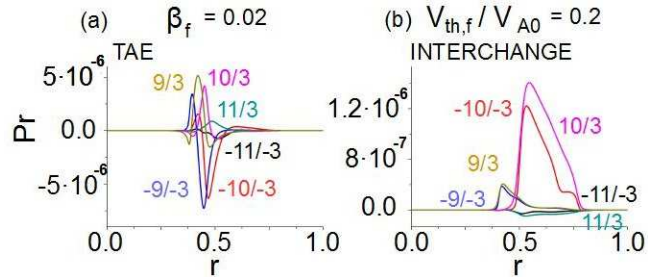}
\caption{Pressure eigenfunction of $n=3$ instability for (a) $\beta_{f}=0.02$ and (b) $V_{th,f}/V_{A0}=0.2$. Case B2b. Each panel includes the instability type.}\label{FIG:14}
\end{figure}

Figure~\ref{FIG:14} shows the pressure eigenfunction of the $n=3$ instability for different $\beta_{f}$ and $V_{th,f}/V_{A0}$ ratios in case B2b. Compared to the eigenfunctions analyzed in Figure~\ref{FIG:14}, if $\beta_{f}=0.02$ $n=3$ instability is a TAE, not a RSAE, panel (a). If $V_{th,f}/V_{A0}=0.2$, the mode is AE stable and an interchange mode is destabilized.

\begin{figure}[h!]
\centering
\includegraphics[width=0.5\textwidth]{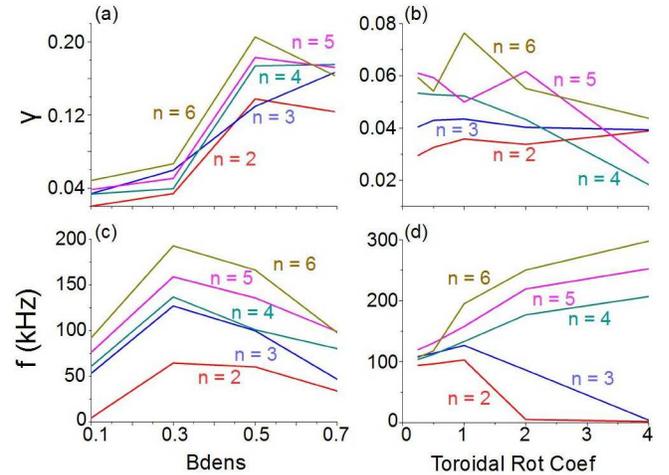}
\caption{Instability growth rate (a) and frequency (c) dependency with the energetic particle density gradient location. Instability growth rate (b) and frequency (d) dependency with the toroidal rotation. Case B2b.}\label{FIG:15}
\end{figure}

\begin{figure}[h!]
\centering
\includegraphics[width=0.5\textwidth]{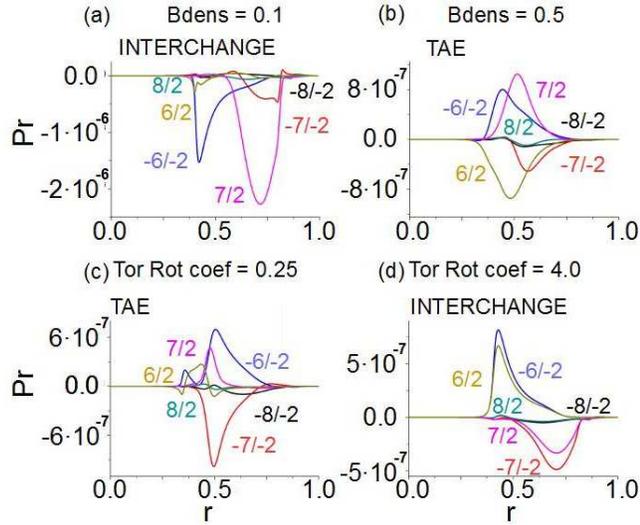}
\caption{Pressure eigenfunction of $n=2$ instability if the energetic density gradient is at (a) $\rho = 0.1$ and (b) $\rho = 0.5$. Instability $n=2$ pressure eigenfunction if the toroidal rotation is (a) $0.25$ times smaller and (b) $4$ times larger compared to the experiment. Case B2b.}\label{FIG:16}
\end{figure}

Figure~\ref{FIG:15} analyzes the effect of the energetic particle density gradient location and the toroidal rotation on the instability's growth rates and frequencies in the reverse shear region. If the energetic particle density gradient is located in the inner plasma ($\rho < 0.3$) the growth rate is minimal, although the largest frequencies are reached if the energetic particle density gradient is at $\rho=0.3$, panels (a) and (c). Consequently, NBI operation optimization requires on axis injection, because $n=2$ is AE stable and $n=3, 4,5,6$ AE show lower growth rates and frequencies. Figure~\ref{FIG:16} shows $n=2$ pressure eigenfunction if the energetic particle density gradient is located nearby the magnetic axis, destabilizing an interchange mode, panel (a), or in the middle plasma where a TAE is triggered, panel (b). A lower toroidal rotation compared to the experiment leads a drop in all mode frequencies, although an increase in the toroidal rotation reduces the frequency of modes $n=2,3$ and an increase in the frequency of $n=4,5,6$, panels (b) and (d) of figure~\ref{FIG:15}. The growth rate of $n=2,3$ is almost the same for the different toroidal rotations analyzed, increasing with the toroidal rotation for $n=4,5,6$. The $n=2$ instability is AE stable and a interchange mode is triggered if the toroidal rotation is 4 times larger than in the experiment, figure~\ref{FIG:16} panel (d), although it is AE unstable and a TAE is triggered if the toroidal rotation is 4 times smaller, panel (c). Consequently, an optimized operation requires plasma rotations $2$ times larger than the experiment to stabilize $n=2$ AE, reduce $n=3$ AE frequency as well as the $n=4,5,6$ growth rates.

\section{Conclusions and discussion \label{sec:conclusions}}

The present study reproduces the most relevant features of the AE activity driven by NBI energetic particles in DIII-D high poloidal $\beta$ discharges at the plasma pedestal, a demonstration of the usefulness of a hybrid reduced MHD/EP Landau closure model for modeling these instabilities. In addition, we have described the main features of the instabilities in discharges with and without bifurcation, identifying the optimal NBI operation regime to suppress or minimize the AE negative effects on plasma stability. The parametric studies performed illustrate the effect of the energetic particle density profile, resonance efficiency, destabilization intensity and plasma toroidal rotation on the AE stability.

The analysis indicates that the bifurcation case is driven if the plasma pedestal ($\rho>0.9$) is decoupled from the rest of the plasma, due to the large thermal ion density and temperature gradient near the periphery, as well as a local maximum of the plasma toroidal rotation near the pedestal. Low $n$ AE are unstable between the middle of the plasma and the periphery (high frequency branch) and $n>3$ ballooning modes are driven at the pedestal (low frequency branch). The bifurcation is triggered if the coupling of poloidal modes of the same toroidal family between the pedestal and the rest of the plasma is weak, evolving independently. On the other hand, in the non bifurcation discharge, the modes of the pedestal and reverse shear region are toroidaly coupled. In addition, observations do not show local maxima of the toroidal rotation in the plasma periphery. Consequently, the pedestal is not isolated and the ballooning modes are stable, although RSAE and TAE/EAE are triggered in the reverse shear region.

To optimize the NBI operation in the bifurcation case during the collapse, NBI injection must be weaker, $\beta_{f} < 0.03$ ($20 \%$ smaller than in the experiment), to keep $n=1,2$ AE stable. The NBI operates in the weak resonance regime if $V_{th,f}/V_{A0} < 0.2$ (this would imply a beam energy almost half of the experimental case). For the bifurcation case after the collapse, NBI injection should be weaker, $\beta_{f} < 0.03$ ($35 \%$ smaller compared to the experiment), to keep $n=1,2,3$ AE stable. If $V_{th,f}/V_{A0} < 0.3$, a beam energy $25 \%$ smaller than the experiment, $n=2,3$ are AE stable and $n=1,4,5,6$ AE show smaller growth rates and frequencies. For the case without bifurcation during the collapse, optimal NBI operation requires lower NBI injection intensity and beam energy, $\beta_{f} < 0.02$ ($45 \%$ smaller) and $V_{th,f}/V_{A0} < 0.2$ (almost half the energy). In summary, NBI operation optimization requires an important modification compared to the experimental operation regime, using weaker injection and beam energy to minimize AE activity. Dedicated DIII-D experiments will be suggested to confirm the optimization trends obtained in the simulations.

\section*{Appendix}

\subsection*{Bifurcation case instability analysis}

\begin{figure*}[h!]
\centering
\includegraphics[width=1.0\textwidth]{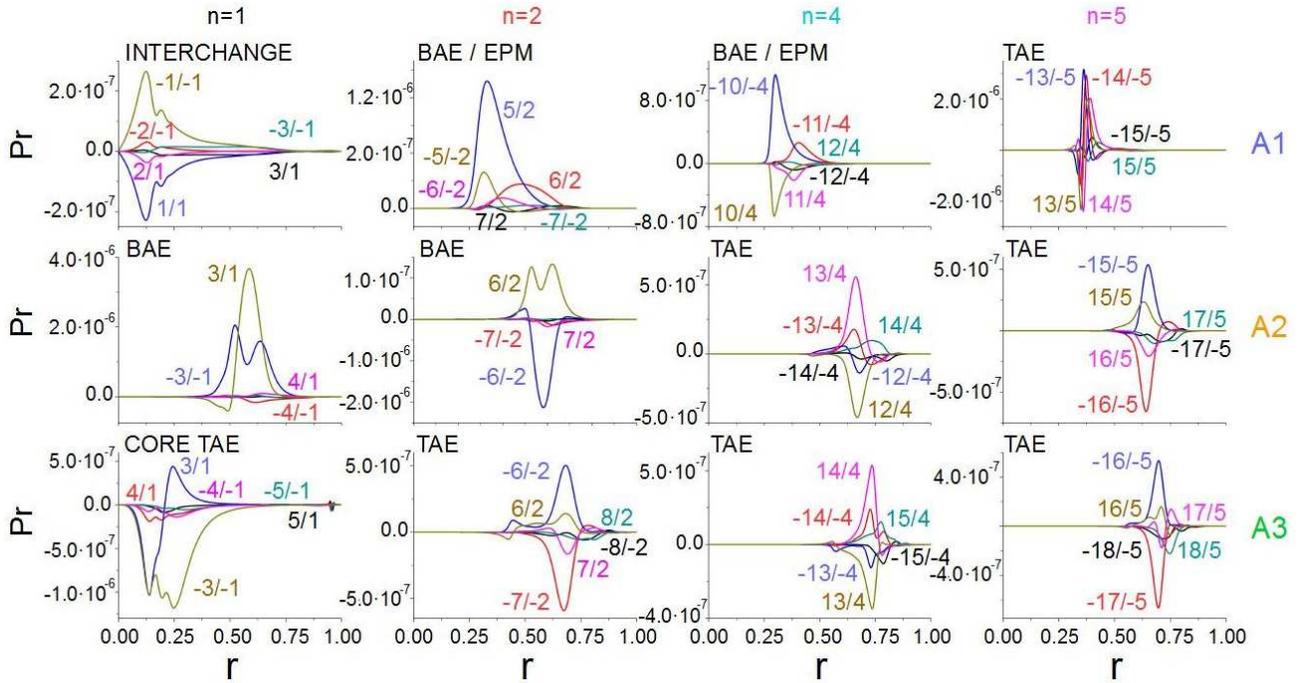}
\caption{Pressure eigenfunctions of the $n=1,2,4,5$ instabilities in the bifurcation case for the discharge phase before (A1), during (A2) and after (A3) the collapse. Each panel includes the instability type.}\label{FIG:17}
\end{figure*}

\begin{figure}[h!]
\centering
\includegraphics[width=0.5\textwidth]{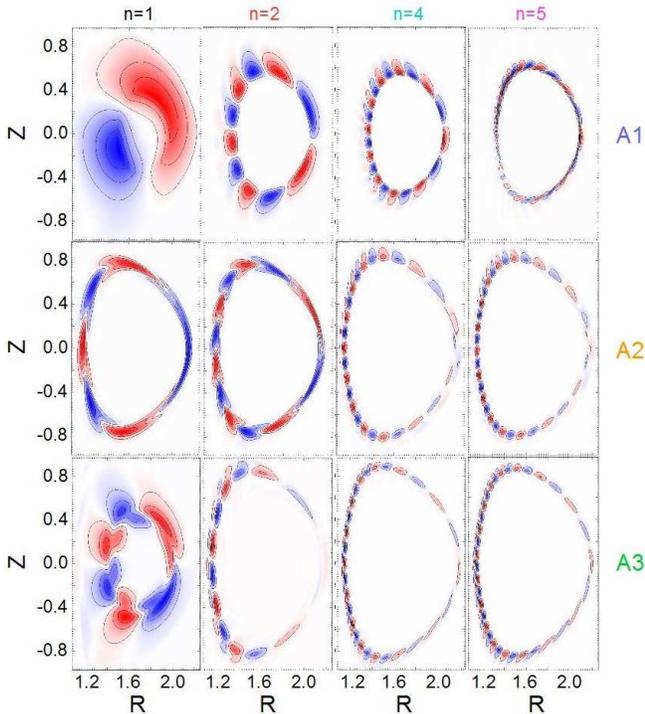}
\caption{2D plots of the $\Phi$ potential for the $n=1,2,4,5$ instabilities in the bifurcation case for the discharge phase before (A1), during (A2) and after (A3) the collapse.}\label{FIG:18}
\end{figure}

\begin{figure*}[h!]
\centering
\includegraphics[width=1.0\textwidth]{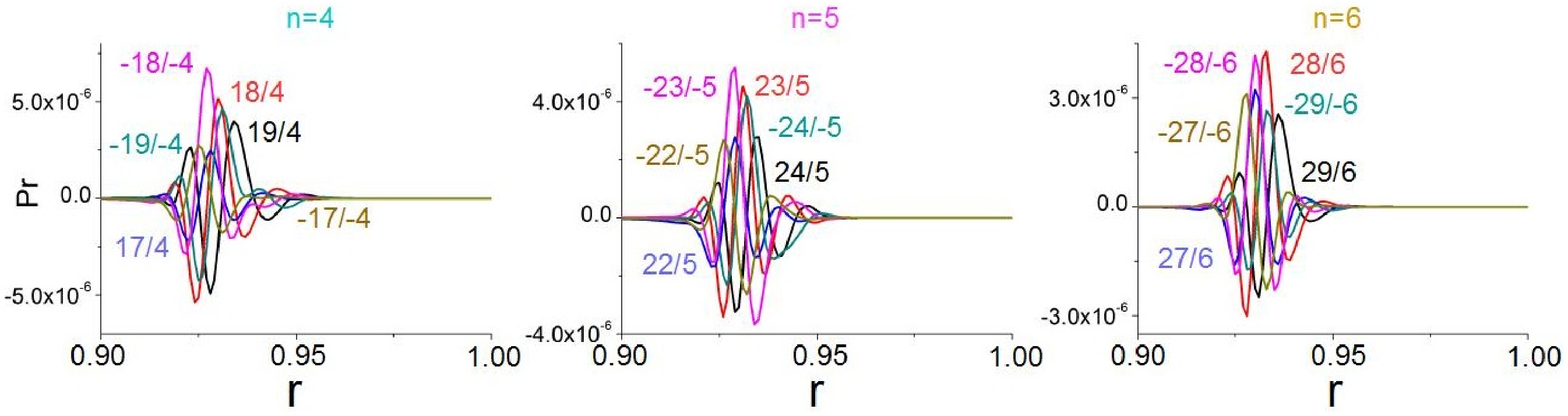}
\caption{Pressure eigenfunctions of the $n=4,5,6$ instabilities in the bifurcation case for the discharge phase after the collapse (simulations limited to the plasma pedestal). Each panel includes the instability type.}\label{FIG:19}
\end{figure*}

Figures~\ref{FIG:17} and \ref{FIG:18} show the pressure eigenfunction and 2D plots of the $\Phi$ potential of $n=1,2,4,5$ instabilities at different discharge phases. Before the collapse, the $n=1$ instability is an interchange mode located close to the magnetic axis because a single mode dominates and there is a weak coupling with the other modes in the plasma. The $n=2$ and $n=4$ are BAE/EPM (Energetic Particle Mode) driven by the modes $5/2$ and $10/4$ because a single mode is dominant, the coupling with other modes is weak and the mode frequency is near the lower continuum gap. The $n=5$ is a TAE because the modes $13/5$ and $14/5$ are coupled and the mode frequency is above $f = 75$ kHz. During the collapse, the $n=1$ evolves into a BAE driven by the modes $3/1$ and the dominant mode of $n=2$ BAE is now $6/2$. The $n=4$ also evolves into a TAE. After the collapse, the $n=1$ is a core TAE, destabilized near the magnetic axis. The other instabilities are TAEs. Figure~\ref{FIG:19} shows the pressure eigenfunction of $n=4,5,6$ instabilities after the collapse if the simulation is limited to the plasma pedestal (A3b). The instabilities are driven around $\rho = 0.925$, showing a stronger toroidal coupling and narrower eigenfunctions than a TAE, pointing out the destabilization of a ballooning mode.

\begin{figure}[h!]
\centering
\includegraphics[width=0.5\textwidth]{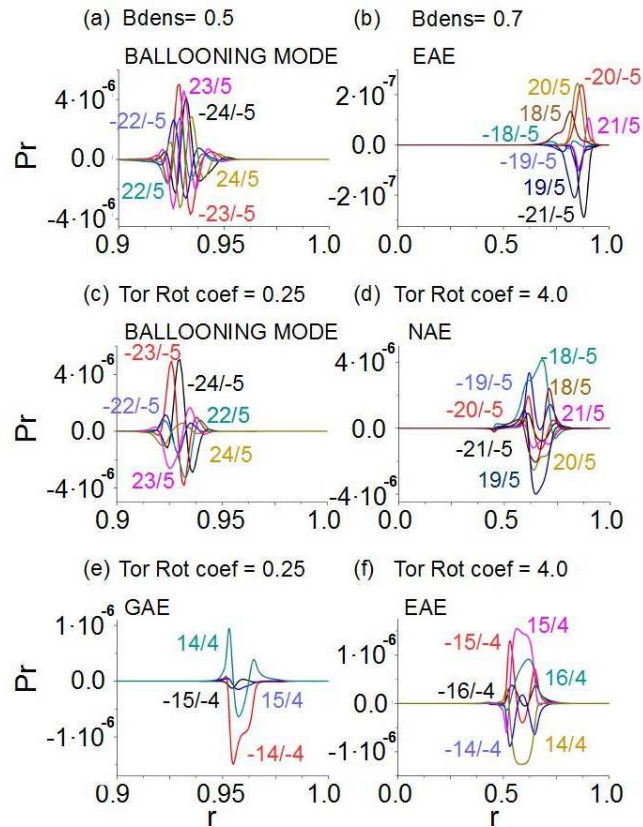}
\caption{Pressure eigenfunctions of $n=5$ instability if the energetic particle density gradient is at (a) $\rho = 0.5$ or (b) $\rho = 0.7$. Pressure eigenfunctions of $n=5$ instability if the toroidal rotation is (a) $0.25$ times smaller and (b) $4$ times larger compared to the experiment. Pressure eigenfunctions of $n=4$ instability if the toroidal rotation is (e) $0.25$ times smaller and (f) $4$ times larger compared to the experiment.}\label{FIG:20}
\end{figure}

Figure~\ref{FIG:20} shows the instability pressure eigenfunction if the energetic particle density gradient is located in the middle of the plasma, panel (a), or in the periphery, panel (b). If the energetic particle density gradient is located in the middle of the plasma, a ballooning mode is destabilized at the pedestal. On the other hand, if the gradient is located in the plasma periphery an EAE is destabilized near the pedestal ($\rho = 0.9$). Reducing the toroidal rotation leads to a drop of the $n=5,6$ instability frequencies, although it increases for $n=4$, because $n=5,6$ instabilities are ballooning modes and $n=4$ is a GAE, panel (e). If the toroidal rotation increases, $n=4$ EAE and $n=5,6$ NAE are destabilized, panels (d) and (f).

\subsection*{Non bifurcation case instability analysis}

\begin{figure*}[h!]
\centering
\includegraphics[width=1.0\textwidth]{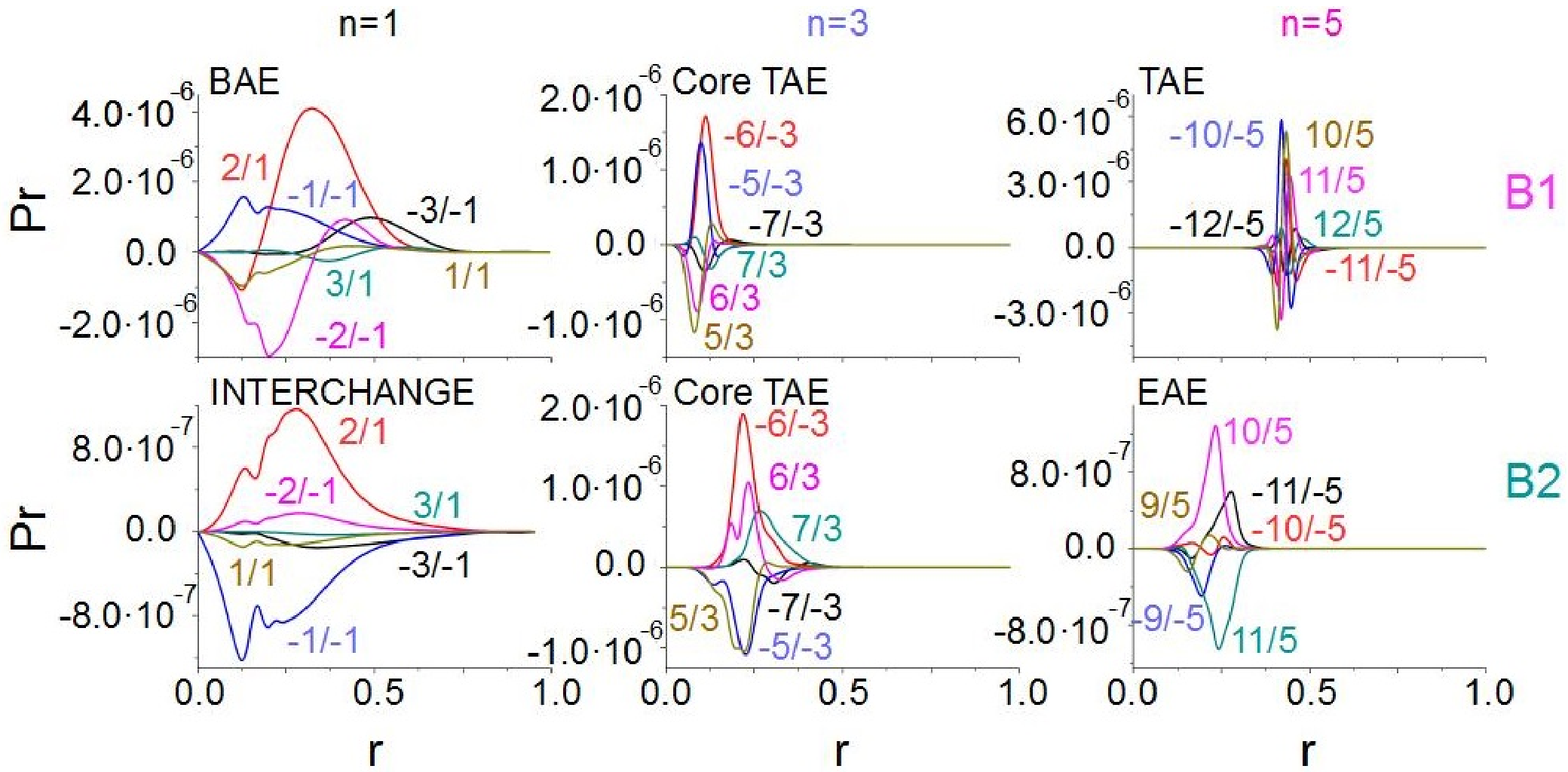}
\caption{Pressure eigenfunctions of $n=1,3,5$ instabilities in the case without bifurcation for the discharge phase before (B1), and after (B2) the collapse. Each panel includes the instability type.}\label{FIG:21}
\end{figure*}

\begin{figure}[h!]
\centering
\includegraphics[width=0.5\textwidth]{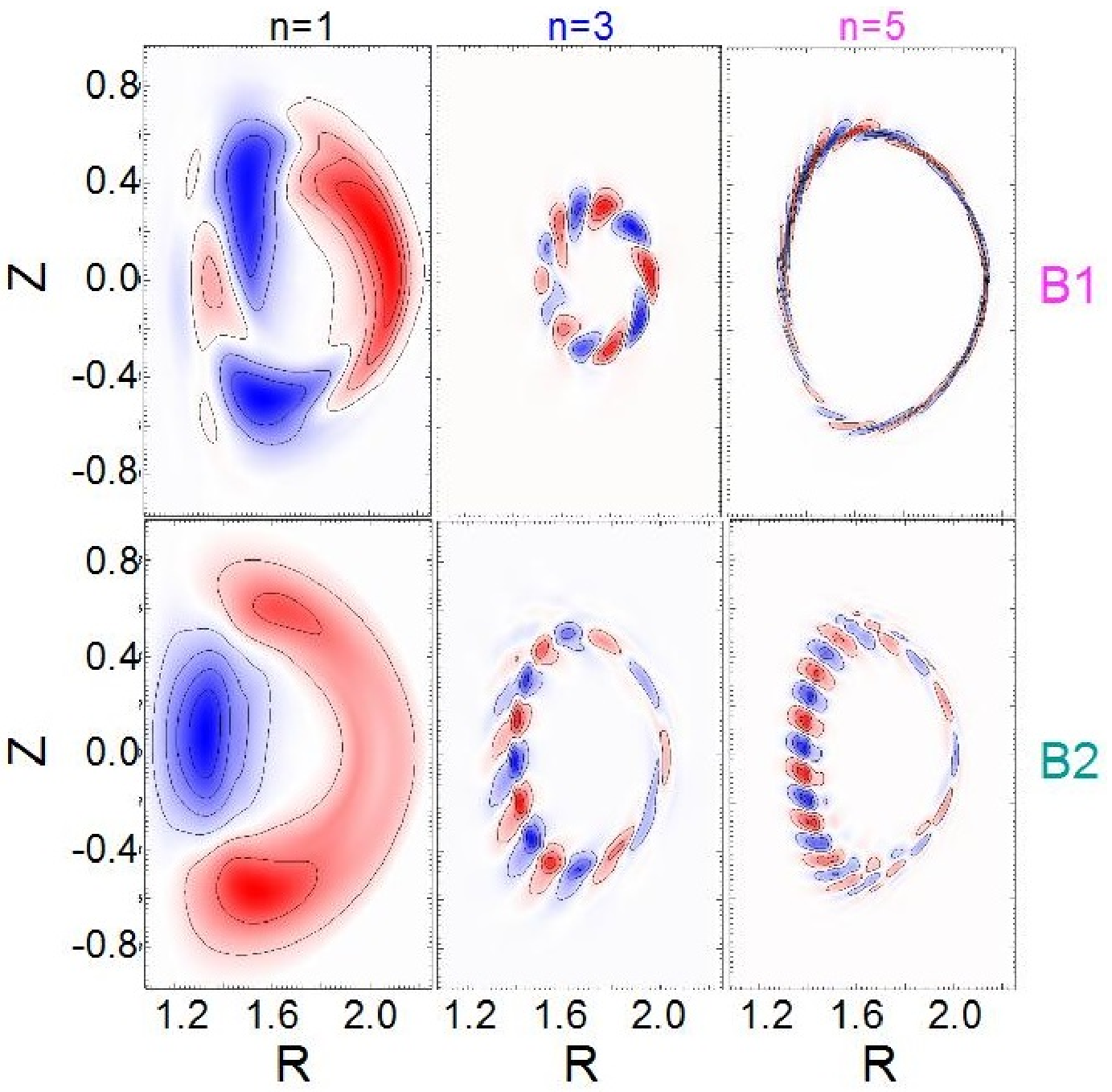}
\caption{2D plots of the $\Phi$ potential for the $n=1,2,4,5$ instabilities in the bifurcation case for the discharge phase before (B1), and after (B2) the collapse.}\label{FIG:22}
\end{figure}

Figures~\ref{FIG:21} and \ref{FIG:22} show the pressure eigenfunction and 2D plots of the $\Phi$ potential of $n=1,3,5$ instabilities at different discharge phases. The $n=1$ instability evolves from a BAE before the collapse to an interchange mode after the collapse. The $n=3$ instability is a core TAE with similar features before and after the collapse, located slightly further from the magnetic axis in B2 phase. The $n=5$ instability evolves from a TAE located in the middle plasma to an EAE destabilized in the inner plasma, because the modes $9/5$ and $11/5$ show a strong coupling and the instability $f$ is $225$ kHz.

\ack
This material based on work is supported both by the U.S. Department of Energy, Office of Science, under Contract DE-AC05-00OR22725 with UT-Battelle, LLC and U.S. Department of Energy, Oﬃce of Science, Oﬃce of Fusion Energy Sciences, using the DIII-D National Fusion Facility, a DOE Oﬃce of Science user facility, under Award No. DE-FC02-04ER54698. DIII-D data shown in this paper can be obtained in digital format by following the links at https://fusion.gat.com/global/D3D\_DMP. This research was sponsored in part by the Ministerio of Economia y Competitividad of Spain under project no. ENE2015-68265-P, National Natural Science Foundation of China Grant No. 11575249, National Magnetic Confinement Fusion Energy Research Program of China under Contract Nos. 2015GB110005, 2015GB102000. The authors also want to acknowledge Prof. W. W. Heidbrink for fruitful discussion.
\\
\\
DISCLAIMER
\\
This report was prepared as an account of work sponsored by an agency of the United States Government. Neither the United States Government nor any agency thereof, nor any of their employees, makes any warranty, express or implied, or assumes any legal liability or responsibility for the accuracy, completeness, or usefulness of any information, apparatus, product, or process disclosed, or represents that its use would not infringe privately owned rights. Reference herein to any specific commercial product, process, or service by trade name, trademark, manufacturer, or otherwise, does not necessarily constitute or imply its endorsement, recommendation, or favoring by the United States Government or any agency thereof. The views and opinions of authors expressed herein do not necessarily state or reflect those of the United States Government or any agency thereof.

\hfill \break

\end{document}